\documentclass[twocolumn,trackchanges,resetfootnote]{aastex7}

\usepackage{hyperref}

\usepackage{siunitx}
\DeclareSIUnit \parsec {pc}
\DeclareSIUnit \arcminSpelled {arcmin}
\DeclareSIUnit \arcsecSpelled {arcsec}
\DeclareSIUnit \erg {erg}

\shorttitle{AGN--Host Galaxy Image Decomposition with JWST}
\shortauthors{Dewsnap et al.}

\received{July 22, 2025}
\revised{September 29, 2025}
\accepted{October 17, 2025}

\submitjournal{PASP}

\graphicspath{{./}{Figures/}}

\begin{document}

\title{AGN--Host Galaxy Image Decomposition with JWST: Limitations of S\'ersic Profile Models}

\author[orcid=0000-0003-1225-3462,gname=Callum,sname=Dewsnap]{Callum Dewsnap}
\affiliation{Department of Physics \& Astronomy, The University of Western Ontario, London, ON, Canada}
\email{cdewsnap@uwo.ca}

\author[orcid=0000-0003-2767-0090,gname=Pauline,sname=Barmby]{Pauline Barmby}
\affiliation{Department of Physics \& Astronomy, The University of Western Ontario, London, ON, Canada}
\affiliation{Institute for Earth \& Space Exploration, The University of Western Ontario, London, ON, Canada}
\email{pbarmby@uwo.ca}

\author[orcid=0000-0001-6217-8101,gname=Sarah C.,sname=Gallagher]{Sarah C. Gallagher}
\affiliation{Department of Physics \& Astronomy, The University of Western Ontario, London, ON, Canada}
\affiliation{Institute for Earth \& Space Exploration, The University of Western Ontario, London, ON, Canada}
\email{sgalla4@uwo.ca}

\correspondingauthor{Callum Dewsnap}
\email{cdewsnap@uwo.ca}

\begin{abstract}

The ability to disentangle the light of an AGN from its host galaxy is strongly dependent on the spatial resolution and depth of the imaging. As the capabilities of imaging systems improve with time, confirming that our standard techniques adequately model the increasingly complex structures unveiled is essential. With JWST providing unprecedented image quality, we can test how measurements of galaxy morphology vary with the choice of point-spread function (PSF) and fitting software. We perform two-component S\'ersic+PSF fits of the surface brightness profiles of 87 X-ray AGNs ($0.1<z<4$) from the CEERS survey. We create model PSFs for NIRCam F115W imaging using both \textsc{photutils} and \textsc{PSFEx}. We find that PSFEx models consistently fail to match the radial profile of typical point sources within our sample. We then perform AGN--host decompositions on each source by creating S\'ersic+PSF models using both \textsc{Galfit} and \textsc{AstroPhot}. We find that \textsc{Galfit} and \textsc{AstroPhot} converge to different regions of the parameter space, providing consistently differing host galaxy properties. While we can measure the AGN and host magnitudes accurately, we find that the host galaxy morphological parameters are not well-determined---the S\'ersic index and effective radius are strongly covariant. Significant changes in the host galaxy parameters do not correspond to changes in the statistical quality of fit, nor to significant changes in the model’s radial profile. These results indicate that the S\'ersic profile does not uniquely well-represent typical AGN host galaxies in extragalactic survey fields. We also provide recommendations for studies of AGN hosts comparable to ours.

\end{abstract}

\keywords{\uat{AGN host galaxies}{2017} --- \uat{Active galaxies}{17} --- \uat{Active galactic nuclei}{16} --- \uat{Galaxies}{573} --- \uat{Galaxy morphology}{582}}

\section{Introduction} \label{sec:intro}

The characteristics of galaxies that contain active galactic nuclei (AGNs) provide valuable insights into the conditions that enable accretion onto supermassive black holes (SMBHs). As a galaxy's structure is closely tied to several evolutionary processes, classifying the morphology is essential for characterizing the host galaxy properties. By investigating sources across a wide redshift range, it is possible to investigate how both the SMBHs and host galaxies evolve over cosmic time. 

When studying high-redshift galaxy morphology, the spatial resolution and sensitivity of the observations are of utmost importance. In the past, such studies relied heavily on observations from the Hubble Space Telescope (HST)---especially for studies of AGN hosts \citep[e.g.]{Haussler2007, Simmons2008, Kim2008, Gabor2009, vanderWel2012, Ding2020, Ji2022, Dewsnap2023}. HST provided the necessary angular resolution and well-characterized point-spread function (PSF) required to disentangle the AGN from the host. However, the relatively low sensitivity and limited reach (in terms of wavelength coverage and field of view) into infrared wavelengths of HST became a substantial limitation on our ability to investigate rest-frame optical morphologies at higher redshifts.

Observations from the James Webb Space Telescope's (JWST) Near Infrared Camera (NIRCam) instrument help alleviate these issues of spatial resolution and sensitivity---NIRCam provides unparalleled spatial resolution and sensitivity, while probing farther into the infrared \citep{Gardner2006, Rigby2023}. This makes it possible to investigate the relationship between AGNs and their host galaxies over a significantly wider cosmic history. These improved JWST observations also bolster the existing studies of galaxy morphology; by providing superior follow-up imaging of existing HST fields, we can reduce uncertainties in previous studies based solely on HST imaging.

The Cosmic Evolution Early Release Science (CEERS) survey \citep{Finkelstein2023} provides an excellent testing ground for studies of AGN host galaxy morphology. CEERS provides imaging and spectroscopy of the Extended Groth Strip (EGS) HST legacy field \citep{Koekemoer2011} using three JWST instruments. Of particular note, CEERS provides ${\sim}100\,\si{\arcminSpelled\squared}$ of NIRCam imaging over seven filters, covering the majority of the EGS field. Due to the coincident EGS field, there exist additional overlapping observations over a wide range of wavelengths. These additional observations allow for the identification of AGNs through X-ray surveys, with optical/IR followup with up to 13 bands of HST and JWST imaging. CEERS has proven to be a valuable dataset for testing how spatial resolution influences the results of galaxy morphology studies, using both parametric and non-parametric methods \citep[e.g.]{Yao2023, Wang2024, Genin2025}.

It is important to compare the techniques we use to separate the light from the AGN from its host. As the depth of modern imaging improves, we find increasing numbers of crowded, overlapping systems that are difficult to de-blend. Furthermore, as more data are collected, we require flexible analysis tools that are capable of handling the sheer amount of data available. Many different image analysis software packages are commonly applied in studies of galaxy morphology, such as \textsc{GIM2D} \citep{Simard1998}, \textsc{Galfit} \citep{Peng2002, Peng2010}, \textsc{BUDDA} \citep{deSouza2004}, \textsc{GASPHOT} \citep{Pignatelli2006}, \textsc{IMFIT} \citep{Erwin2015}, \textsc{galight} \citep{Ding2021}, or \textsc{AstroPhot} \citep{Stone2023}. There is significant overlap in the use cases and implementation for each software package, and so understanding the relationships between their final outputs is essential for evaluating their efficacy.

To this end, we investigate the significance of the choice of PSF on the morphological fits of a source by applying two common PSF-building tools, \textsc{PSFEx} \citep{Bertin2011} and \textsc{photutils} \citep{Bradley2022} and comparing the results. We investigate how the fits differ for two common image-fitting software packages, \textsc{Galfit} \citep{Peng2002, Peng2010} and \textsc{AstroPhot} \citep{Stone2023}. We apply our techniques to both JWST and HST data to determine how significantly the results differ on AGN hosts between $0.1<z<4$.

The structure of the paper is as follows. In Section \ref{sec:data}, we outline the AGN selection criteria and the imaging used in our fits. Section \ref{sec:psfs} outlines the selection process for our point source library and the process of generating our PSFs. Section \ref{sec:fits} outlines the fitting process used for both image-fitting software packages. We review the results of our morphological fits in Section \ref{sec:results}. In Section \ref{sec:discussion}, we evaluate the differences between the resulting \textsc{AstroPhot} and \textsc{Galfit} fit parameters, as well as discuss how best to interpret the S\'ersic profile. We outline the main takeaways and provides recommendations for best practices moving forward in Section \ref{sec:conclusion}.

All magnitudes throughout use the AB system \citep{Oke1983}. We adopt a flat $\Lambda$CDM cosmology with $H_0=70\,\si{\kilo\metre\per\second\per\mega\parsec}$, $\Omega_\mathrm{m}=0.3$, and $\Omega_\mathrm{\Lambda}=0.7$.

\section{Data} \label{sec:data}

\subsection{Imaging} \label{subsec:imaging}

CEERS is one of the early release science surveys of JWST. The NIRCam imaging uses the F115W, F150W, F200W, F277W, F356W, F410M, and F444W filters with $5\sigma$ point-source limiting magnitudes varying from 28.2--29.2 \citep{Finkelstein2023}. The data are given in 10 mosaics, split between two data releases. The first data release (DR0.5) fields were observed in June 2022 and included four pointings, CEERS1--3 and CEERS6. The second data release (DR0.6) fields were observed roughly six months later in December 2022 and included the remaining six pointings (CEERS4, CEERS5, CEERS7--10). The DR0.6 fields were observed with a \SI{180}{\degree} flip relative to DR0.5. We use the final mosaics provided by the CEERS team\footnote{\url{https://ceers.github.io/releases.html}}; see \citet{Bagley2023} for details on the data reduction including a discussion regarding their background subtraction process. The mosaics in the seven filters are all pixel-aligned with pixel scales of 0.03''/pix. Additionally, the CEERS team provides pixel-aligned images for all available HST filters for each pointing. These filters are F606W and F814W for HST/ACS (Advanced Camera for Surveys) and F105W, F125W, F140W, and F160W for HST/WFC3 (Wide Field Camera 3).

In this study, we focus primarily on the NIRCam F115W filter. This specific band was selected for a number of reasons: firstly, this filter probes the rest-frame optical for the majority of sources in our sample (see Section \ref{subsec:agn} and Figure \ref{fig:L_x_vs_z} for a description of our sample). Thus, our host galaxy measurements probe the stellar population while minimizing dust extinction. While including a higher fraction of the stellar population makes distinguishing between the AGN and the host more difficult, this is counterbalanced by the depth and resolution provided by the F115W observations. F115W provides the deepest point-source limiting magnitude among the CEERS JWST observations; specifically, the $5\sigma$ limiting AB magnitude is 29.2. In addition, the resolution of the imaging is the best of all JWST filters---\citet{Finkelstein2023} list the FWHM as \SI{0.066}{\arcsec}.

As a comparison throughout, we also apply a subset of our methodology to imaging from HST's F814W filter. This imaging was selected as F814W is ubiquitous throughout similar studies both prior to and post JWST coming online. The majority of the arguments for using JWST's F115W filter also apply here. Although F606W does provide deeper imaging than F814W for the EGS field ($5\sigma$ point source limiting magnitude of 28.6 compared to 28.3) at a slightly higher resolution (FWHM of \SI{0.118}{\arcsec} versus \SI{0.124}{\arcsec}), F606W probes a significantly different population than F115W. On the other hand, any HST imaging at wavelengths longer than F814W see a significant drop-off in resolution and depth. The nearest filter in terms of wavelength would be either F105W or F125W, which provide significantly brighter limiting magnitudes of 27.1 and 27.3 and FWHM values of \SI{0.235}{\arcsec} and \SI{0.244}{\arcsec}, respectively \citep{Finkelstein2023}. F814W provides the best balance of resolution, depth (necessary for the AGN--host decomposition), and wavelength (necessary due to the wavelength dependence of the S\'ersic index) to compare with F115W.

\subsection{AGN Catalog} \label{subsec:agn}

We select our sample of AGNs using the AEGIS-XD survey \citep{Nandra2015}, which imaged the central EGS region with \textit{Chandra}. We use X-ray selected AGNs due to the robustness of the identification. While X-ray surveys do not provide a complete sample of AGNs, there is little contamination from inactive galaxies \citep{Civano2012}. Other surveys, such as optical or infrared, can be contaminated by objects such as star-forming galaxies and provide a less reliable sample \citep{Donley2008, Donley2012}.

\begin{figure}[t]
\includegraphics[width=0.46\textwidth]{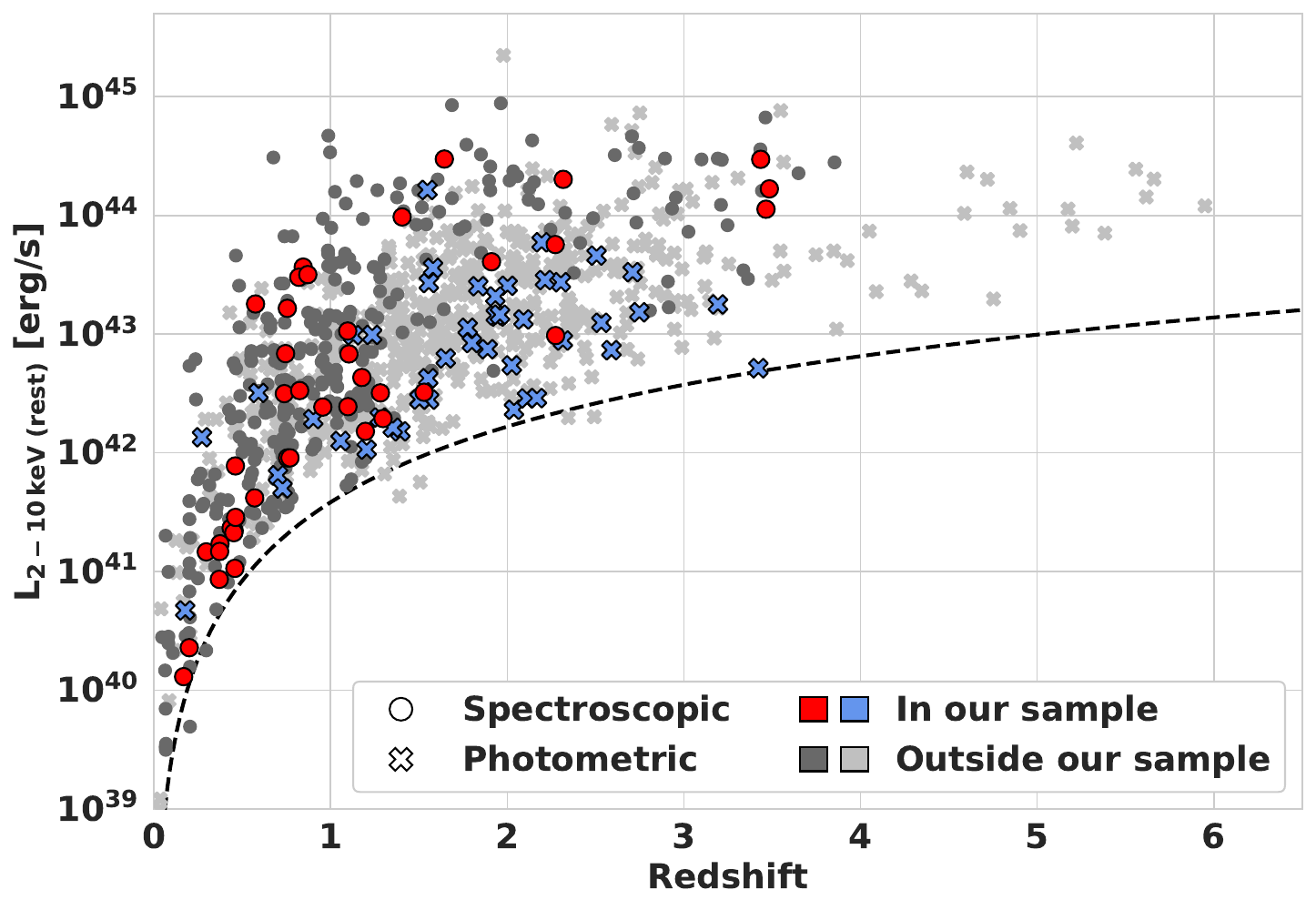}
\caption{The rest-frame X-ray luminosity in the \SIrange{2}{10}{\kilo\electronvolt} band from \citet{Nandra2015} as a function of redshift. Sources in red and blue are included in our sample, whereas the gray and dark gray sources are not contained in our sample. Round markers are sources with a confident spectroscopic redshift, whereas crosses indicate photometric redshifts from multiband SED fitting of up to 35 bands of UV to mid-IR imaging. The dashed line represents the X-ray flux limit converted to luminosity. The X-ray luminosities have been K-corrected to rest-frame 2--10~keV.}
\label{fig:L_x_vs_z}
\end{figure}

\citet{Nandra2015} provides a catalog of 937 X-ray point sources, alongside multiband counterparts from the Rainbow Cosmological Surveys Database \citep{Barro2011a, Barro2011b} for 929 of the 937. Of these sources, 353 have reliable spectroscopic redshifts from Keck and MMT, with the remainder having redshifts based on multiband photometry from the observed-frame UV to mid-IR. The limiting X--ray flux of this survey is $1.5\times10^{-16} \ \si{\erg\per\centi\metre\squared\per\second}$ in the full band (0.5--10 keV), corresponding to $L_X = 3.80 \times 10^{38} \ \si{\erg\per\second}$ to $L_X = 2.31 \times 10^{43} \ \si{\erg\per\second}$ at $z=0.1$ and $z=4$, respectively.  Figure~\ref{fig:L_x_vs_z} shows the X-ray luminosity versus redshift of the sample.  Following \citet{Nandra2015}, we used the observed \SIrange{0.5}{10}{\kilo\electronvolt} flux to calculate the rest-frame \SIrange{2}{10}{\kilo\electronvolt} luminosity, applying a K-correction with an assumed power-law X-ray spectrum with $\Gamma=1.4$, assuming the flux is unabsorbed. 

Using this catalog, we find the X-ray sources which fall within the bounds of the CEERS NIRCam pointings. This accounts for 121 of the 937 sources. We apply \textsc{SExtractor} \citep{Bertin1996} to both the JWST and HST imaging, and search for the nearest detected source to the position of the multiband counterpart listed in the catalog. Since the JWST and HST images are aligned, the position of the counterpart is fixed between the two sets of images. After all sources are detected, we visually inspect that each source selected as the counterpart in our imaging is correct, inspecting all 13 bands provided by the CEERS team.

During the \textsc{SExtractor} step, we excluded eight sources from our sample which had no detectable counterpart near the position provided by \citet{Nandra2015}. Additionally, we cut another 16 sources where a significant portion of the source is cut off by the edge of the imaging or a chip gap. Lastly, a number of sources lie within a region of low signal-to-noise in the mosaic image due to the observing pattern. This is particularly common in CEERS5 and CEERS7--9 due to the additional F115W imaging obtained during CEERS' wide field slitless spectroscopy (WFSS) observations included in the mosaic. This results in another 10 sources being removed from our sample, leaving us with a total of 87. Of our 87 sources, 41 have spectroscopic redshifts above the confidence threshold employed by \citet{Nandra2015}, leaving 46 remaining sources with photometrically determined redshifts. See Figure \ref{fig:L_x_vs_z} for the X-ray luminosity of our sources as a function of redshift. From here, we create a cutout for each source for both JWST and HST on which we perform our fits.

Occasionally, a source will appear in multiple pointings due to small regions of overlap. In these cases, we create our cutout from the pointing in which the source is closest to the center of the detector. The overlap is typically in the regions of low signal-to-noise (due to the observing and dithering patterns), and as such choosing between the two is straightforward.

\section{Point-Spread Functions} \label{sec:psfs}

\subsection{PSF Construction} \label{subsec:construction}

While both \textsc{photutils} and \textsc{PSFEx} generate PSF models by stacking images of individual point sources within an image, \citet{Zhuang2024} find that the characteristics of the generated PSFs differ significantly. They report that, in general, \textsc{PSFEx} outperforms \textsc{photutils} in AGN--host decompositions. Their study primarily focused on a field much more densely packed with point-source candidates, leading to orders of magnitudes more point-source candidates than are available in the CEERS field. They did however generate PSFs for the CEERS field as well, and performed an example fit on a source within our sample. See Section \ref{subsec:characteristics} for a comparison of our generated PSFs and Section \ref{sec:discussion} for a comparison of the example source (the broad-line quasar SDSS1420+5300A). Notably, \citet{Zhuang2024} were only able to investigate the DR0.5 data, as the DR0.6 data had not yet been made publicly available.

\begin{figure}[t]
    \centering
    \includegraphics[width=\linewidth]{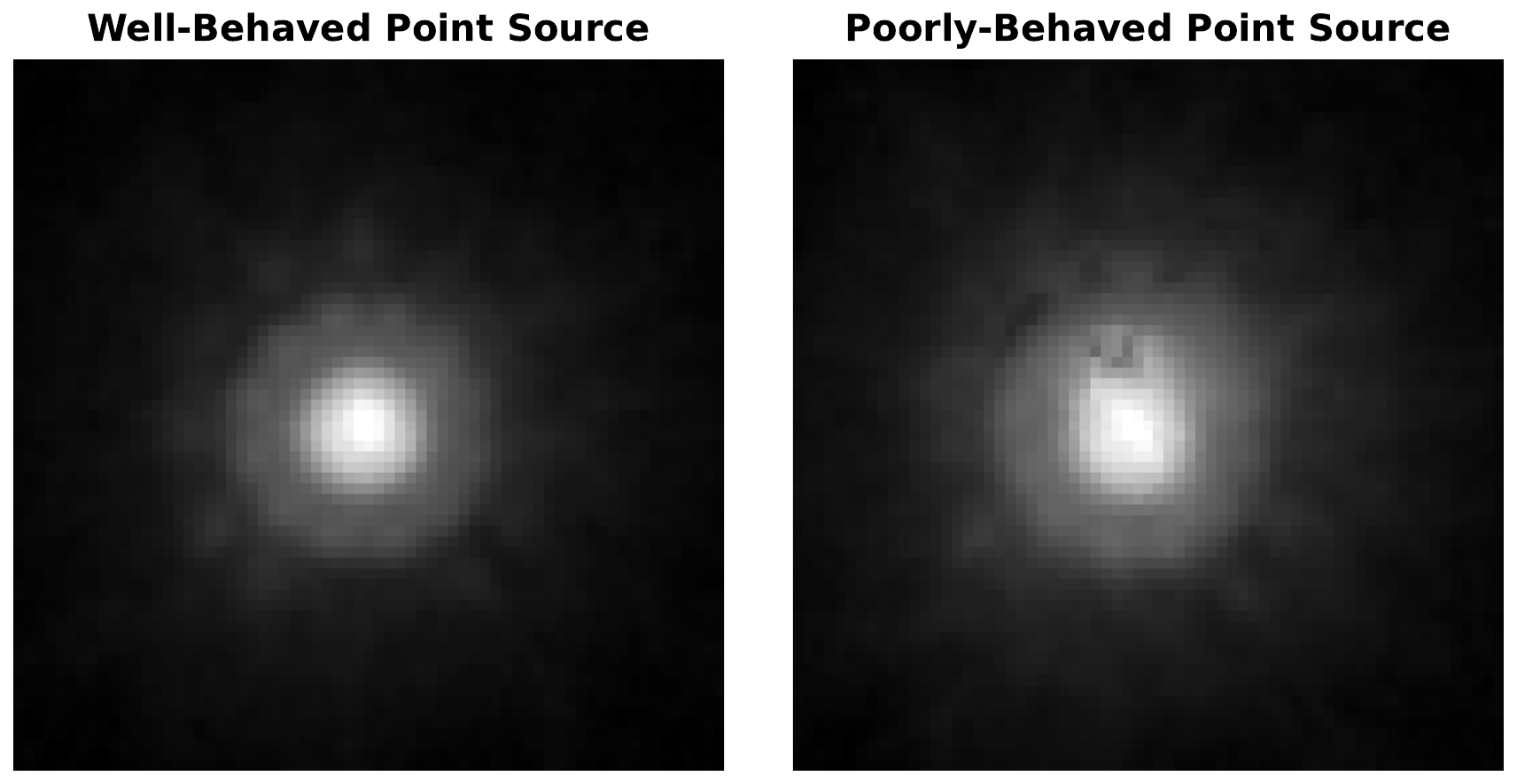}
    \caption{Example of a poorly behaved HST/ACS F814W point-source candidate (right) alongside a well-behaved candidate (left). Each cutout shows the central $2\si{\arcsec}\times2\si{\arcsec}$ region of the point source. The poorly behaved point source shows several anomalous pixels on the upper region of the core; these effects can significantly contaminate any PSF model constructed using the point source.}
    \label{fig:bad_star}
\end{figure}

In creating our point source catalog, we followed a selection process and analysis similar to \citet{Zhuang2024}. We apply \textsc{SExtractor} to detect and select candidate point sources in our field. In particular, we searched for sources that are likely to be stellar ($\texttt{CLASS\_STAR} \geq 0.8$), have high signal-to-noise ratio ($\texttt{SNR\_WIN} \geq 20$), have sufficiently low FWHM ($\texttt{FWHM}_\mathrm{JWST} \leq 3$ pixels and $\texttt{FWHM}_\mathrm{HST} \leq 6$ pixels), are not blended ($\texttt{FLAGS} \leq 2$), and are not significantly elongated ($\texttt{ELONGATION} < 1.5$). Following this preliminary filtering, we manually search through our sample and remove any spurious sources. This includes sources too close to the edge of the detectors, saturated sources, or any blended sources not caught by the \texttt{FLAGS} check. A significant number of the highest signal-to-noise HST candidates, while not saturated, were significantly disfigured by substantial azimuthal asymmetry (see Figure \ref{fig:bad_star} for an example). Many such sources passed the \textsc{SExtractor} criteria (although worse examples were flagged)---asymmetric  sources significantly influence the parameters of the generated PSF if left in the sample and may require individual visual inspection to identify and discard. Thus, a significant number of high signal-to-noise candidates are dropped from our HST catalog. We additionally ensure that none of our point-source candidates overlap with any point-like sources within our AGN sample.

This search provides us with a total of 89 and 64 point-source candidates for JWST and HST, respectively. These are the sources that we feed into our PSF-building software. For each source, we generate a cutout of 135$\times$135 pixels (4.05\si{\arcsec}$\times$4.05\si{\arcsec}) for JWST. This value was selected as it contains ${\sim}98$\% of the light contained within the PSF \citep{Rigby2023}. Similarly, we generate a 167$\times$167 pixel (5.01\si{\arcsec}$\times$5.01\si{\arcsec}) cutout for each HST point source. The HST cutout size is due to the broader HST PSF. Note that the DR0.6 observations were taken with JWST rotated \SI{180}{\degree} relative to the DR0.5 observations; due to this, we orient all our point source cutouts to match the DR0.5 orientation such that we can compare between epochs most accurately throughout Section \ref{sec:psfs}. See \citet{Bagley2023} for the positioning of the JWST modules and detectors relative to the DR0.5 orientation and the JWST documentation\footnote{\url{https://jwst-docs.stsci.edu/jwst-near-infrared-camera}} for a breakdown of the specific modules and detectors.

We begin constructing our PSFs by first arranging our JWST point sources into a variety of groupings. The goal of these groups is to investigate how significantly the choice of point sources influences the model generated for each PSF-building software package. The groups are defined by their data release, module, detector, dithering pattern, and/or signal-to-noise ratio. The majority of these groupings will not be used in the fits described throughout Sections \ref{sec:fits}--\ref{sec:discussion}; they exist only to compare the characteristics of the output PSF models. See Section \ref{sec:app_groups} for a complete description of the groups.

We begin creating our PSFs by passing each of our groups into \textsc{PSFEx}. The majority of \textsc{PSFEx} parameters are left as default. We applied a \texttt{PIXEL\_AUTO} basis type with a $\texttt{BASIS\_NUMBER} = 10$, $\texttt{SAMPLE\_VARIABLILITY} = 0.5$, and a $\texttt{SAMPLE\_FWHMRANGE}_\mathrm{JWST}=1-4$ pixels (0.03\si{\arcsec}--0.12\si{\arcsec}) or $\texttt{SAMPLE\_FWHMRANGE}_\mathrm{HST}=2-8$ pixels (0.06\si{\arcsec}--0.24\si{\arcsec}). For both \textsc{PSFEx} and \textsc{photutils}, we generate a model oversampled by a factor of 2, with a final model size of 265$\times$265 pixels (3.975\si{\arcsec}$\times$3.975\si{\arcsec}) for JWST or 325$\times$325 pixels (4.875\si{\arcsec}$\times$4.875\si{\arcsec}) for HST. Due to the limited number of point sources within our region, we do not utilize the spatial variation function within \textsc{PSFEx}. We only create constant PSFs; the subgroupings will act as our most ``local'' PSFs.

We also pass the final sample of selected stars to \textsc{photutils}, with the parameter \texttt{maxiters} varying from 2--10 based on the number of point sources in the sample. For both \textsc{PSFEx} and \textsc{photutils}, we attempted multiple runs with varied input parameters and point-source groupings. The PSFs generated by \textsc{PSFEx} were fairly robust to the choice of input settings, whereas \textsc{photutils} models saw significant dependence on the choice of \texttt{maxiters} for groups with a smaller number of point sources.

\subsection{PSF Characteristics} \label{subsec:characteristics}

\begin{figure*}[t]
    \centering
    \includegraphics[width=\linewidth]{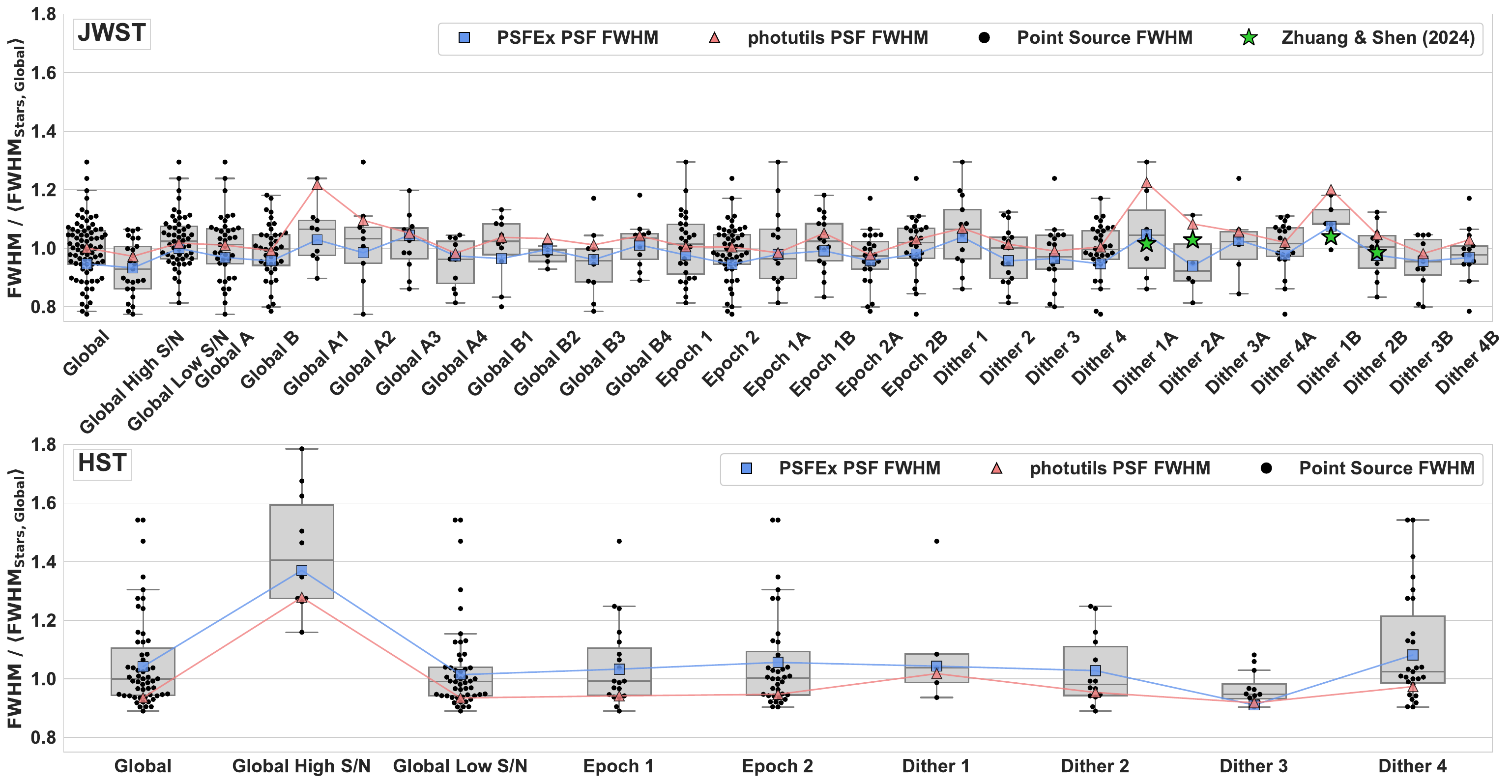}
    \caption{The distribution of the FWHM of each PSF model and their constituent point-source sample for JWST (top panel) and HST (bottom panel) as measured by \textsc{AstroPhot}. The FWHM for each is normalized by the median FWHM of the Global point source sample. The black points represent each point source within the sample. The gray box-and-whisker plots provide information on the distribution of the point sources: the central line in the box gives the median value, the extent of the box gives the quartiles above and below, and the whiskers include the rest of the sample. The red and blue points represent the PSF models for \textsc{photutils} and \textsc{PSFEx}, respectively. The green stars are the PSF FWHM values from \citet{Zhuang2024}; only four of our JWST groupings overlap their study.}
    \label{fig:FWHM}
\end{figure*}

In order to get a consistent measure of the properties of each PSF, we model the central core of the PSF with a Gaussian. For comparison with \citet{Zhuang2024}, we similarly define the central region as $2N+1$, where $N$ is the smallest integer number of pixels larger than half the FWHM. We assume an initial estimate of the FWHM to be those reported by \citet{Finkelstein2023}, namely $\mathrm{FWHM_{F115W}}=\SI{0.066}{\arcsec}$ and $\mathrm{FWHM_{F814W}}=\SI{0.124}{\arcsec}$. We perform point-source model fits with both \textsc{Galfit} and \textsc{AstroPhot}, finding consistent results with both. Additionally, we also analogously model every point source within each group.

Figure \ref{fig:FWHM} summarizes the FWHM for each of our PSF models for both JWST and HST alongside their constituent point sources as measured by \textsc{AstroPhot}. Looking first at JWST, we find, in agreement with \citet{Zhuang2024}, that the FWHM of the \textsc{photutils} PSF models is consistently higher than those of the \textsc{PSFEx} models. However, each PSF's placement relative to the distribution of the point sources included in the sample changes from group to group. For groups with a significant number of point sources (e.g., Global, Global Low S/N, Epoch 2, etc.), we find that the \textsc{photutils} PSF is very similar to the median FWHM of the group's point sources, whereas the \textsc{PSFEx} PSF tends to be closer to the bottom quartile of the boxplot. When we look at groups with a low number of point sources (e.g., Global A1, Dither 1A, Dither 1B, etc.), both \textsc{photutils} and \textsc{PSFEx} tend to have a higher FWHM; particularly \textsc{photutils}, where the measured FWHM occasionally escapes the upper quartile of the point-source population. In these cases, \textsc{PSFEx} tends to provide a FWHM closer to the global point-source median.

We additionally overlay the \citet{Zhuang2024} CEERS PSFs, another set of \textsc{PSFEx}-generated models. The \citet{Zhuang2024} PSFs tend to behave somewhat similarly to our \textsc{PSFEx} PSFs, although the sample size is low in all four overlapping groups. This leads to high variance in the measured FWHMs, as well as significant dependence on the choice of specific point sources within the grouping across all three models. Our models also provide the axis ratio ($b/a$) of the core, where we find consistently high measures of axial symmetry ($\geq$0.97), again in agreement with \citet{Zhuang2024}.

For our HST measurements, we find very consistent FWHMs for both \textsc{photutils} and \textsc{PSFEx}, except for one clear outlier: the Global High S/N group. This group possesses a significantly higher FWHM measurement, likely due to more undetected poorly behaved point sources, as discussed in Section \ref{subsec:construction} and Figure \ref{fig:bad_star}. When these spurious point sources are included in other groups, they are often rejected by the PSF-fitting software, as they are significant outliers when accompanied by a larger sample of more typical point sources. Interestingly, we find that \textsc{photutils} PSFs tend to have a consistently lower FWHM than \textsc{PSFEx}, opposite to what we see in the JWST PSFs. With significant numbers of quality point sources, the two PSFs tend to straddle the global point source median, with \textsc{PSFEx} just above and \textsc{photutils} just below. The consistency is much higher from group to group for HST, although that is expected---the groupings are defined by JWST observing parameters.

\begin{figure*}[t]
    \centering
    \includegraphics[width=\linewidth]{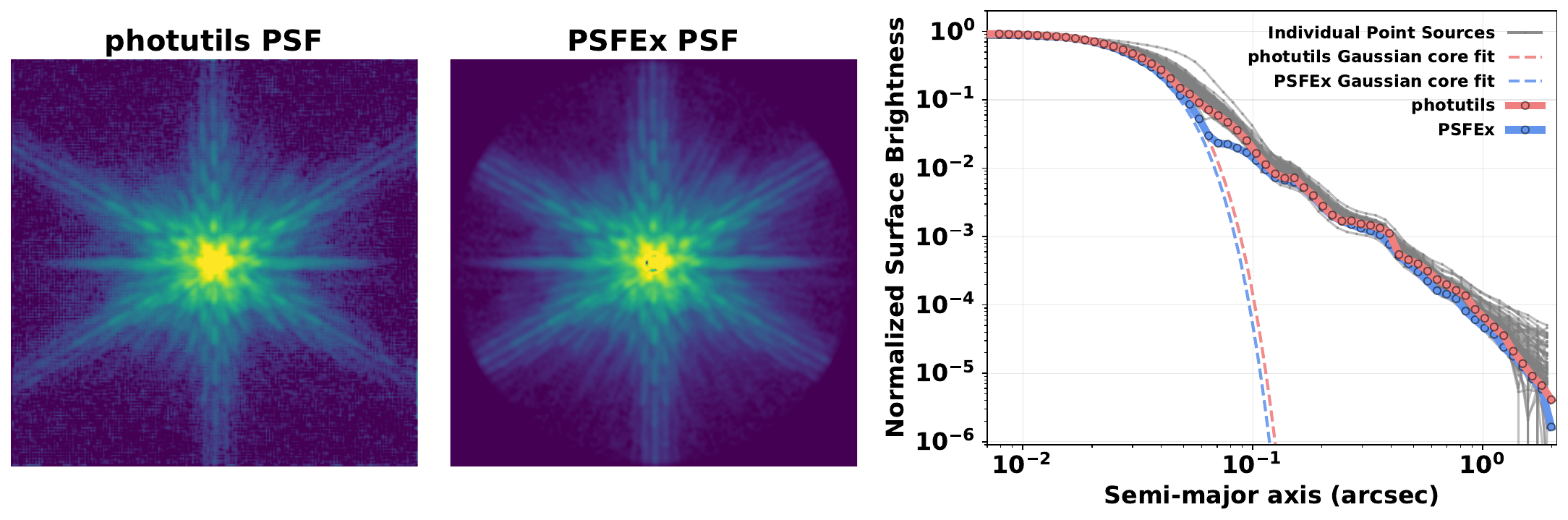}
    \caption{An example of a \textsc{photutils} (left) and \textsc{PSFEx} (middle) PSF for JWST/NIRCam's F115W filter. Each cutout is 4\si{\arcsec} in diameter. This particular example features the Global grouping. The rightmost plot shows the 1D radial profile of each PSF (\textsc{photutils} in red, \textsc{PSFEx} in blue) alongside their constituent point sources in gray. The dashed lines represent the Gaussian fit of the PSF core we used to determine the PSF FWHM.}
    \label{fig:1D_PSF}
\end{figure*}

\begin{figure*}[t]
    \centering
    \includegraphics[width=\linewidth]{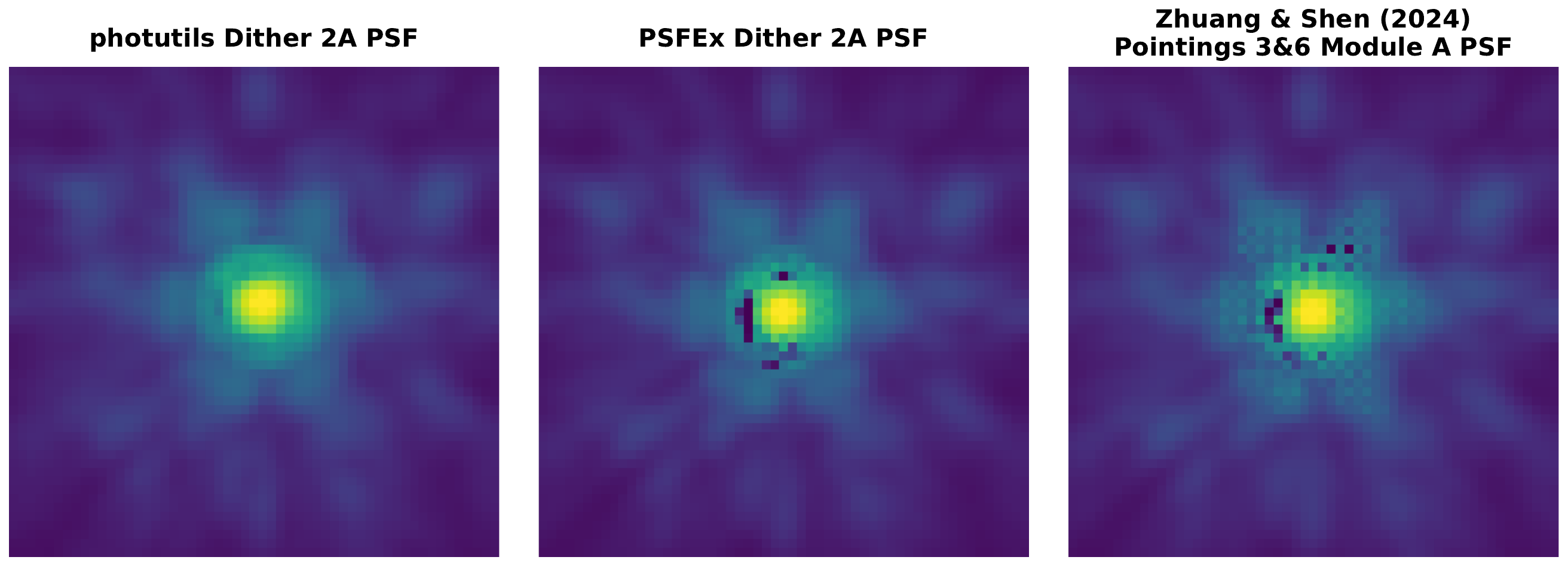}
    \caption{Comparison of the central \SI{0.825}{\arcsec}$\times$\SI{0.825}{\arcsec} region of the Dither 2A grouping's PSFs for JWST/NIRCam F115W. The \textsc{photutils} and \textsc{PSFEx} models are the left and middle panels, respectively. We also include the publicly available \textsc{PSFEx} model generated by \citet{Zhuang2024} in the right panel. All three models are normalized to the same log scale.}
    \label{fig:core_comparison}
\end{figure*}

Next, we investigate the radial profile of our PSF models. We apply \textsc{photutils} isophotal fitting to each of our PSFs and point sources. Figure \ref{fig:1D_PSF} shows an example of such a fit, specifically for the JWST Global grouping. Both PSFs, as well as the point sources, generally possess a very similar radial profile. However, the \textsc{PSFEx} PSF has a clear divot away from the rest of the profiles just outside of its FWHM, at roughly \SIrange{0.06}{0.1}{\arcsec}. This feature is present to some degree in all of our JWST \textsc{PSFEx} PSFs, but in none of our \textsc{photutils} PSFs. No point sources in our entire sample demonstrate this divot. Looking towards our HST models, we do not see this stark difference between the two PSFs. The center of the \textsc{PSFEx} PSF in Figure \ref{fig:1D_PSF}, shows some artifacts occurring a few pixels away from the center of the model. Figure \ref{fig:core_comparison} shows a zoomed-in view of this central region for another of our models, namely the Dither 2A grouping, alongside one of the \citet{Zhuang2024} PSFs. We can clearly see anomalous pixels at the edge of the central core in similar regions of both \textsc{PSFEx} models, while there are no such effects in the \textsc{photutils} PSF. Due to the presence of these irregularities at $r\sim\mathrm{FWHM}$, it is likely a cause of the consistently lower FWHM we see in JWST's \textsc{PSFEx} PSFs.

In general, both PSF software provide consistent measurements of the extended wings of the PSF, and the visual construction is very clear and consistent. However, the wings of the \textsc{photutils} PSF quickly degrade when \texttt{maxiters} is set too high for the number of point sources in the sample, whereas \textsc{PSFEx} produces more consistent results. When the number of sources in the sample is low ($<$10), noise begins to dominate at the outer reaches of the \textsc{photutils} PSF model after only a few iterations, while the structure of the \textsc{PSFEx} PSF is relatively well-defined. On the other hand, \textsc{photutils} never encounters the issue of irregularities in the radial profile, and produces closer matches between the typical point source profile across all groups.

\section{Fitting Process} \label{sec:fits}

Now that we have identified our AGN candidates and constructed PSF models, we can proceed to fitting our sources. We apply both \textsc{Galfit} and \textsc{AstroPhot} in order to compare the results between two often used 2D photometric fitting software packages. We follow the same process for both; we use a two-component model, with a S\'ersic profile representing the host galaxy and a point-source model representing the AGN. By fitting the S\'ersic and point-source components simultaneously, we can effectively disentangle the light from each source with minimal contamination. If the light from the central AGN is left unaccounted for, light is misattributed to the bulge, thus artificially inflating the S\'ersic index and contaminating the characterization of morphology along with any galaxy magnitude or size measurement. Two-component fitting such as this is standard and has been thoroughly tested across both 1D and 2D-fitting \citep{Haussler2007, Simmons2008}.

The S\'ersic profile \citep{Sersic1963, Sersic1968} is defined as
\begin{equation}
    \Sigma (r) = \Sigma_e \exp{\left[ -\kappa_n \left( \left( \frac{r}{r_e} \right)^{1/n} -1 \right) \right]},
\end{equation}
with $\Sigma_e$ the pixel surface brightness at the half-light radius $r_e$. The half-light radius is defined as the radius of the isophote containing half of the luminosity of the galaxy. The parameter $n$ is the S\'ersic index, and $\kappa_n$ is a variable dependent on $n$ defined by
\begin{equation}
    \gamma (2n,\kappa_n) = \frac{1}{2} \Gamma (2n),
\end{equation}
where $\Gamma$ and $\gamma$ are the Gamma and lower incomplete Gamma functions, respectively.

The S\'ersic index is especially significant, as it acts as proxy for a morphological classification. For example, $n=1/2$ corresponds to a Gaussian, $n=1$ provides us with the exponential disk model typically applied to disk-dominated galaxies \citep{Freeman1970}, and $n=4$ gives us the de Vaucouleurs law, often applied to bulge-dominated sources \citep{deVaucouleurs1948, deVaucouleurs1976, deVaucouleurs1991}. Conveniently, if one integrates the S\'ersic profile out to $r=\infty$ to calculate the total brightness of a source, there exists an analytic solution given by
\begin{equation} \label{eq:flux}
    F_\mathrm{total} = 2 \pi r_e^2 \Sigma_e \mathrm{e}^{\kappa_n} n \kappa_n^{-2n} \Gamma(2n) q,
\end{equation}
where $q=b/a$ is the axis ratio.

\textsc{Galfit} and \textsc{AstroPhot} use a very similar set of free parameters by default and, in practice, we run a very similar fitting process for each. We start by taking the cutouts for each of our sources as defined in Section \ref{subsec:agn}. We apply \textsc{SExtractor} to each cutout in order to obtain a set of initial parameters, as well as to identify any neighbors that we need to address. We use a segmentation map generated by \textsc{SExtractor} to create our mask by visually inspecting each source and determining which neighbors may significantly influence our fit. If we determine that a neighbor is not significant, we simply mask it out. If it is significant, we add an additional S\'ersic profile to our fit to model the neighboring source.

For our main source (the AGN+host) and neighbors, we use the parameters from the \textsc{SExtractor} catalog as our preliminary initial conditions. These parameters include the positions of the components, combined magnitude of the AGN and its host, the effective galaxy radius, the galaxy axis ratio, and the position angle. We then manually adjust the initial conditions via visual inspection to ensure that our model is fitting all sources as needed and has high quality initial conditions. These manual adjustments include varying the initial S\'ersic index, the AGN-to-host flux ratio, dealing with any over-deblended (e.g., features such as a spiral arm deblended into a separate source) or under-deblended sources (e.g., overlapping neighbor not detected as distinct from the main source), etc. Both \textsc{Galfit} and \textsc{AstroPhot} allow for the user to apply constraints to the parameters of the fit. These constraints are selected such that the parameters remain physical, while not overly limiting variation during the fitting process---in order to find the true solution, the parameters may need to vary beyond typically expected values. We select a similar set of constraints to \citet{Dewsnap2023}, summarized in Table \ref{tab:constraints}. We enforce no explicit constraints regarding the total flux of the source, nor on how the light is distributed between the host galaxy and AGN components---both are allowed to vary independently.

 Next, we used \textsc{AstroPhot} with the Global \textsc{photutils} PSF for an initial fit to refine the initial parameter values. The final result of our fits are not sensitive to choice of fitting software or model PSF for this initial fit. If the fit is deemed reasonable by visual inspection, we move on with the fit results as our final set of initial conditions---if not, we further adjust the initial parameters by hand and iterate through the fitting process again.

\begin{deluxetable}{lc}
\tablecaption{The parameter constraints applied to each fit in \textsc{Galfit} and \textsc{AstroPhot}. \label{tab:constraints}}
\tablehead{
\colhead{Parameter Name} & \colhead{S\'ersic+PSF Fit}}
\startdata
    Host S\'ersic Index $\left(n\right)$ & $0.36<n<8$\\
    Host Axis Ratio $\left(q\right)$ & $0.01<q<1$\\
    $x_\mathrm{host}$ Position [arcsec] & $\left|x_\mathrm{host} - x_\mathrm{init}\right| \leq 0.15\si{\arcsec}$ \\
    $y_\mathrm{host}$ Position [arcsec] & $\left|y_\mathrm{host} - y_\mathrm{init}\right| \leq 0.15\si{\arcsec}$\\
    $x_\mathrm{AGN}$ Position [arcsec] & $\left|x_\mathrm{AGN} - x_\mathrm{init}\right| \leq 0.15\si{\arcsec}$\\
    $y_\mathrm{AGN}$ Position [arcsec] & $\left|y_\mathrm{AGN} - y_\mathrm{init}\right| \leq 0.15\si{\arcsec}$\\
\enddata
\tablecomments{The parameter constraints listed for the host are also applied to any neighboring sources.}
\end{deluxetable}

Once we have our reasonable initial fit values, we proceed to running fits using both \textsc{Galfit} and \textsc{AstroPhot} with a wide range of PSFs. In particular, we use the Global, Global A/B, Epoch 1A--2B, and Dither A1--B4 groups for our JWST fits, leading to 16 fits per source (2 fitting software packages $\times$ 2 PSF software $\times$ 4 PSF groupings). These groupings were selected as they are representative of the PSF distribution; they contain sets with both high and low numbers of point sources, global and local sample selections, and a wide range of FWHM measurements. This allows for a robust investigation of dependence on choice of PSF of the fit results while limiting the study to a reasonable number of fits per source.

\section{Results} \label{sec:results}

\begin{figure*}
    \centering
    \includegraphics[width=\linewidth]{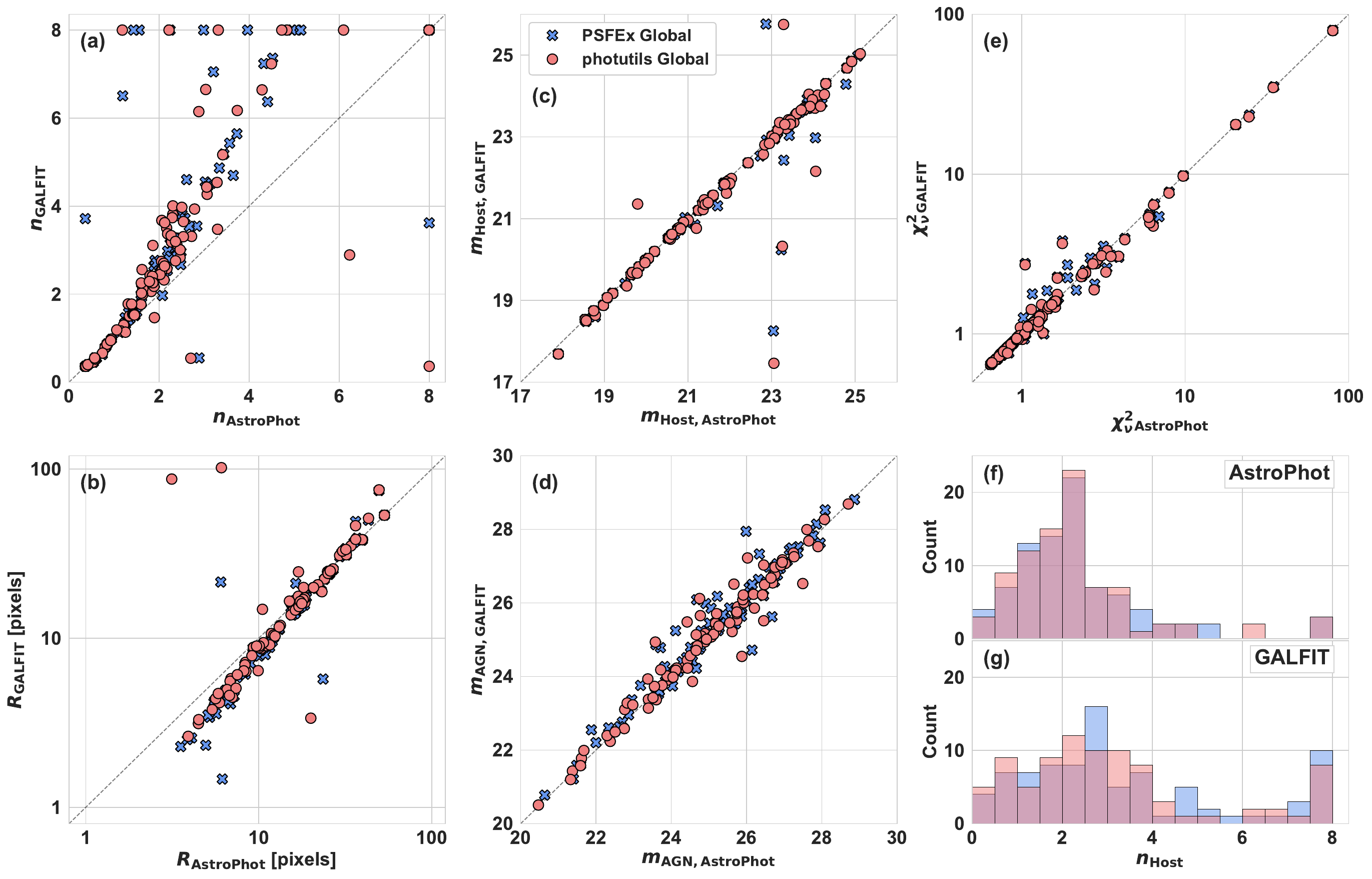}
    \caption{Comparison of the \textsc{Galfit} and \textsc{AstroPhot} fits using the Global PSF for both \textsc{PSFEx} (blue crosses) and \textsc{photutils} (red circles). The dashed line represents one-to-one agreement. The parameters included are: (a) the S\'ersic index, (b) effective radius, (c) host magnitude, (d) AGN magnitude, (e) $\chi^2_\nu$, and the distribution of the S\'ersic index for \textsc{AstroPhot} (f) and \textsc{Galfit} (g). We find a clear disagreement between the S\'ersic index and effective radius measurements between \textsc{Galfit} and \textsc{AstroPhot}. We also note that a few of the most significant outliers fall outside the bounds of the plots.}
    \label{fig:AstroPhot_vs_GALFIT}
\end{figure*}

Figure \ref{fig:AstroPhot_vs_GALFIT} shows a direct comparison of the \textsc{AstroPhot} and \textsc{Galfit} fit results. For this comparison, we show only the fits which use the Global groups for each software package. The Global grouping is representative of the whole sample---see Section \ref{sec:app_fits} for a discussion on the robustness of the fit results between the PSF groupings and software. We find that the effective radii of the host galaxies typically fall within \SIrange{0.1}{3}{\arcsec}, with host magnitudes between \SIrange{25}{18}{mag}. The AGN components have a fainter distribution, falling between \SIrange{29}{20}{mag}.

The most striking feature of Figure \ref{fig:AstroPhot_vs_GALFIT} is the strong deviation from one-to-one agreement seen in both the S\'ersic index (Figure \ref{fig:AstroPhot_vs_GALFIT}a) and the effective radius (Figure \ref{fig:AstroPhot_vs_GALFIT}b). The distribution of the S\'ersic indices (Figure \ref{fig:AstroPhot_vs_GALFIT}f and \ref{fig:AstroPhot_vs_GALFIT}g) show that at $n\geq2$, \textsc{Galfit} tends to find significantly higher values of $n$, often reaching the upper boundary of $n\sim8$. The distribution of sources between $0.5<n<4$ is also relatively flat. \textsc{AstroPhot} produces far fewer fits with high values of $n$, peaking between $1<n<2.5$. This shift in $n$ is counterbalanced by the effective radius also shifting away from one-to-one, specifically at lower radii ($r_e<10$ pixels).

Interestingly, the host magnitude (Figure \ref{fig:AstroPhot_vs_GALFIT}c) remains consistent despite the stark difference between the other host galaxy parameters. The scatter within the AGN magnitude (Figure \ref{fig:AstroPhot_vs_GALFIT}d) and $\chi^2_\nu$ (Figure \ref{fig:AstroPhot_vs_GALFIT}e) also shows no correlation with the sources that strongly disagree in $n$ and $r_e$. In general, the variation of the AGN magnitude is of similar order between the choice of PSF and the choice of fitting software, however the choice due to the fitting software is more symmetrical about zero.

We note that any additional attempts to account for the background, such as adding a plane sky model in either \textsc{Galfit} or \textsc{AstroPhot}, does not significantly affect our fit results, nor help further constrain the S\'ersic index and effective radius. In general, our sources are generally sufficiently bright ($m_\mathrm{Host} < 23$) such that the contribution of the sky is low relative to the flux of the source itself. For the fainter sources, the choice of PSF and fitting software plays a more significant role on the results, leading the systematic uncertainties to dominate those of the sky subtraction. Since the influence of the sky subtraction is on the extended wings of the galaxy. However, the $\chi^2_\nu$ weighs the wings much less than the higher S/N regions of the source, meaning that small changes in the wings do not lead to significant changes in the best fit models.

Overall, the S\'ersic index and effective radius depend most strongly on the choice of fitting software, whereas the AGN magnitude depends more significantly on the choice of PSF, as expected. The host magnitude and $\chi^2_\nu$ are consistent throughout each combination of PSF and fitting software, after accounting for clear outliers.

\section{Discussion} \label{sec:discussion}

\begin{figure*}
    \centering
    \includegraphics[width=\linewidth]{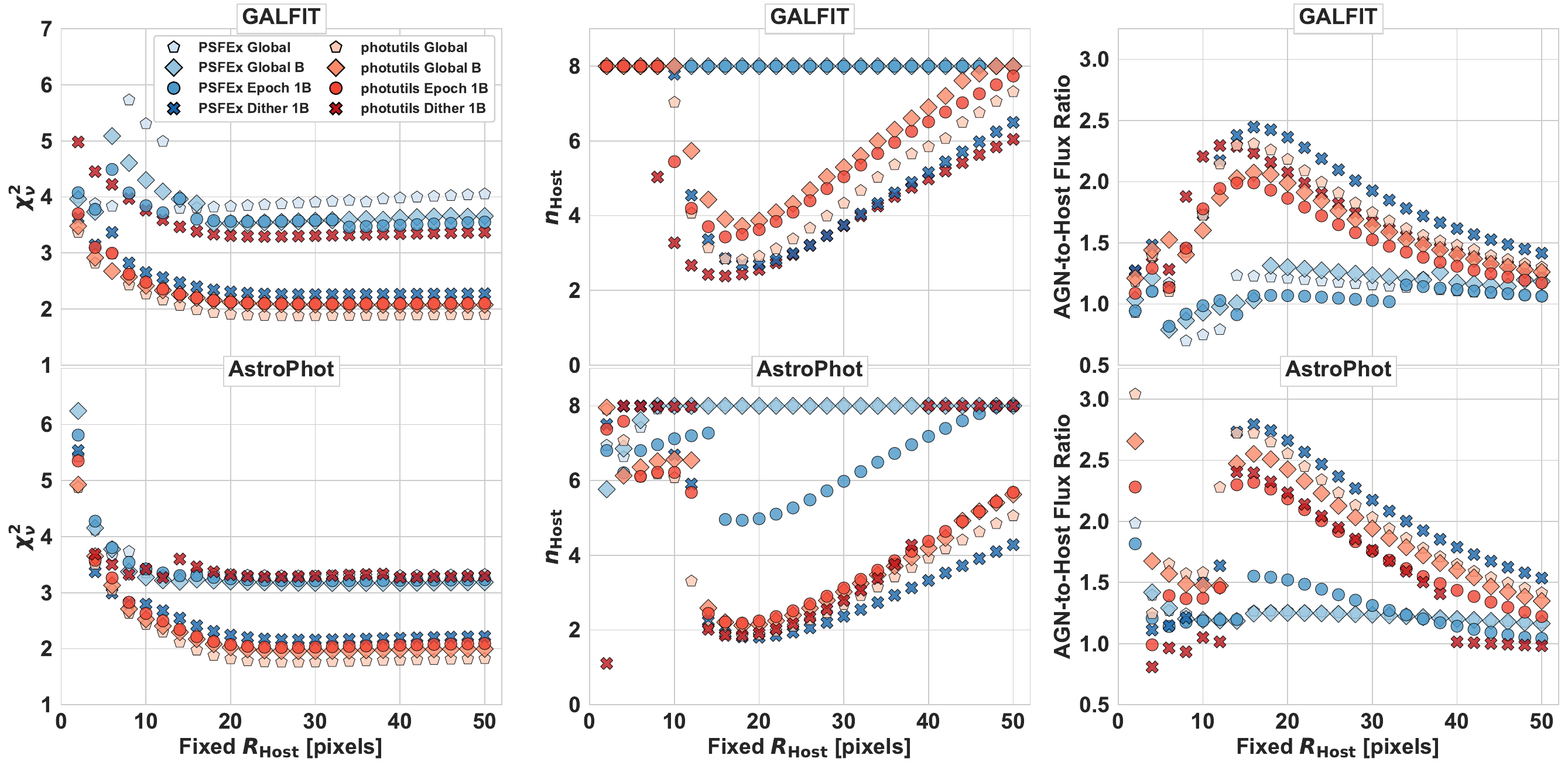}
    \caption{Comparison of the parameter space probed by our \textsc{Galfit} (top row) and \textsc{AstroPhot} (bottom row) fits. Each marker represents a fit to the 2-D surface brightness of SDSS1420+5300A with the effective radius held fixed at the given value, while every other parameter freely varies. The blue and red markers correspond to fits using \textsc{PSFEx} and \textsc{photutils} PSFs, respectively. The marker style represents the point source grouping of each PSF.}
    \label{fig:Degen_RedBlue}
\end{figure*}

The strong disagreement of the S\'ersic index and effective radius between \textsc{AstroPhot} and \textsc{Galfit} while the host magnitude and $\chi^2_\nu$ remain consistent suggests that there may exist degenerate solutions for measurements of the host galaxy. It is not necessarily the case that one software package is performing better than the other; it is possible that they are simply parsing different regions of the parameter space and resulting in two equally “good” best fits, but exist in consistently different regimes.

The possibility that the effective radius and S\'ersic index cannot be individually well-determined is not a novel concept. \citet{Ji2020, Ji2022} investigate the covariance of the effective radius and S\'ersic index, finding that the S\'ersic index is consistently not well-constrained, even for high signal-to-noise (${\sim}100$) sources. Similarly, \citet{Dewsnap2023} performed an in-depth analysis of the parameter space of high-redshift ($0.03<z<6.5$) AGN hosts imaged by HST, finding a strong, positive correlation between S\'ersic index and effective radius.

To investigate the parameter space our fits exist in, we focus on a small subset of sources on which we perform a wide range of fits. We define an array of values for the effective radius over which we perform a set of fits with the effective radius held fixed throughout the minimization process. Then, we can determine how the various parameters interact through a wider region of the parameter space.

In this section, we specifically highlight the source SDSS1420+5300A (Rainbow ID: AEGIS 742, RA: 215.02339290, DEC: +53.01020700), a broad-line quasar at $z_\mathrm{spec}=1.644$ with $M_\mathrm{BH} = 10^{9.00\pm0.16} M_\sun$ \citep{Grier2019}. This source has been extensively studied, providing an excellent example with which we can compare the results of our fits with those throughout the literature, and is representative of the results seen throughout our sources. Decomposing SDSS1420+5300A reveals a massive ($M_*=10^{11.3}M_\sun$) barred spiral galaxy, oriented face-on. \citet{Ding2022} perform several brightness profile fits using \textsc{galight} \citep{Ding2021}, leading to a spatially resolved 2D SED fit of the outer regions (0.24”) of the host. However, \citet{Ding2022} do not perform fits with the F115W imaging, and thus we can only compare our results qualitatively. \citet{Zhuang2024} also perform several surface brightness fits of this source using \textsc{Galfit}, providing an excellent example to compare directly with our fits.

We begin by forming an array of effective radius values ranging over 2--50 pixels (0.06--1.5\si{\arcsec}). For each value, we perform a fit with the effective radius held fixed at the corresponding value. We use the same initial conditions for remaining parameters as described in Section \ref{sec:fits}; however, we do not fit any neighboring sources to minimize any possible differences from fit to fit (although we still apply our masks). All the parameters except for the effective radius are free to vary within the constraints outlined in Table \ref{tab:constraints}. We perform each fit using both \textsc{AstroPhot} and \textsc{Galfit} for each PSF model. For SDSS1420+5300A, this corresponds to the Global, Global B, Epoch 1B, and Dither 1B groupings.

Figure \ref{fig:Degen_RedBlue} shows the $\chi^2_\nu$, S\'ersic index, and AGN-to-host flux ratio as a function of effective radius for each of the models generated. For both \textsc{AstroPhot} and \textsc{Galfit}, the fits fall into two clear groups: the ones that successfully separate the AGN from the host galaxy, and those that do not. The successful fits are characterized by a few key elements. First, the $\chi^2_\nu$ is consistently lower, indicating a higher quality of fit. Next, the fits result in lower S\'ersic indices, whereas the failed fits result in much higher indices, usually $n\sim8$. Because we know by both visual inspection and through previous studies that the host galaxy is a barred spiral, intuitively we expect that the S\'ersic index should be roughly $1<n<3$. Lastly, the successful fits find a consistently higher AGN-to-host flux ratio. This is expected, as the source is a known luminous quasar and, as such, the AGN dominates the host. When the fits fail, it is because the minimization algorithm fails to identify the host galaxy beneath the AGN. Thus, the S\'ersic and point-source components tend to share the AGN light, resulting in a more point-like S\'ersic profile with an AGN-to-host ratio ${\sim}1$.

Generally, the fits that perform well in one fitting software also fit well for the other---for this source, the poorer fits are preferentially the \textsc{PSFEx} PSFs; specifically, the Global, Global B, and Epoch 1B PSFs. However, the Dither 1B \textsc{PSFEx} PSF does perform well. Notably, the Dither 1B PSF shows far less influence by the anomalous pixels discussed in Section \ref{sec:psfs} compared to the other three \textsc{PSFEx} models. Since SDSS1420+5300A is so AGN-dominated, the anomalous pixels have a larger influence on the resulting fits, likely explaining why the \textsc{PSFEx} fits perform disproportionately worse than the \textsc{photutils} fits.

One interesting result is that the \textsc{photutils} Dither 1B PSF fits are grouped with the failed fits for the $\chi^2_\nu$, but are more closely aligned with the successful fits in terms of the measured host parameters. This is likely because the \textsc{photutils} Dither 1B is among the poorest models of the \textsc{photutils} PSFs, with one of the largest FWHM values of all our PSF models. As such, it is likely that the higher $\chi^2_\nu$ is due in large part to the poor PSF. Of note is the \textsc{AstroPhot} \textsc{photutils} Dither 1B fits specifically---once the host-galaxy parameters begin to align with the rest of the good fits, the $\chi^2_\nu$ increases slightly. Then, at $r_e\geq40$ pixels, the model begins to fail again and reaches $n\sim8$, where we see the $\chi^2_\nu$ decrease. This is counterintuitive—one would expect that if the model were more reasonably fitting the host that the $\chi^2_\nu$ would decrease, not increase.

\begin{figure*}
    \centering
    \includegraphics[width=\linewidth]{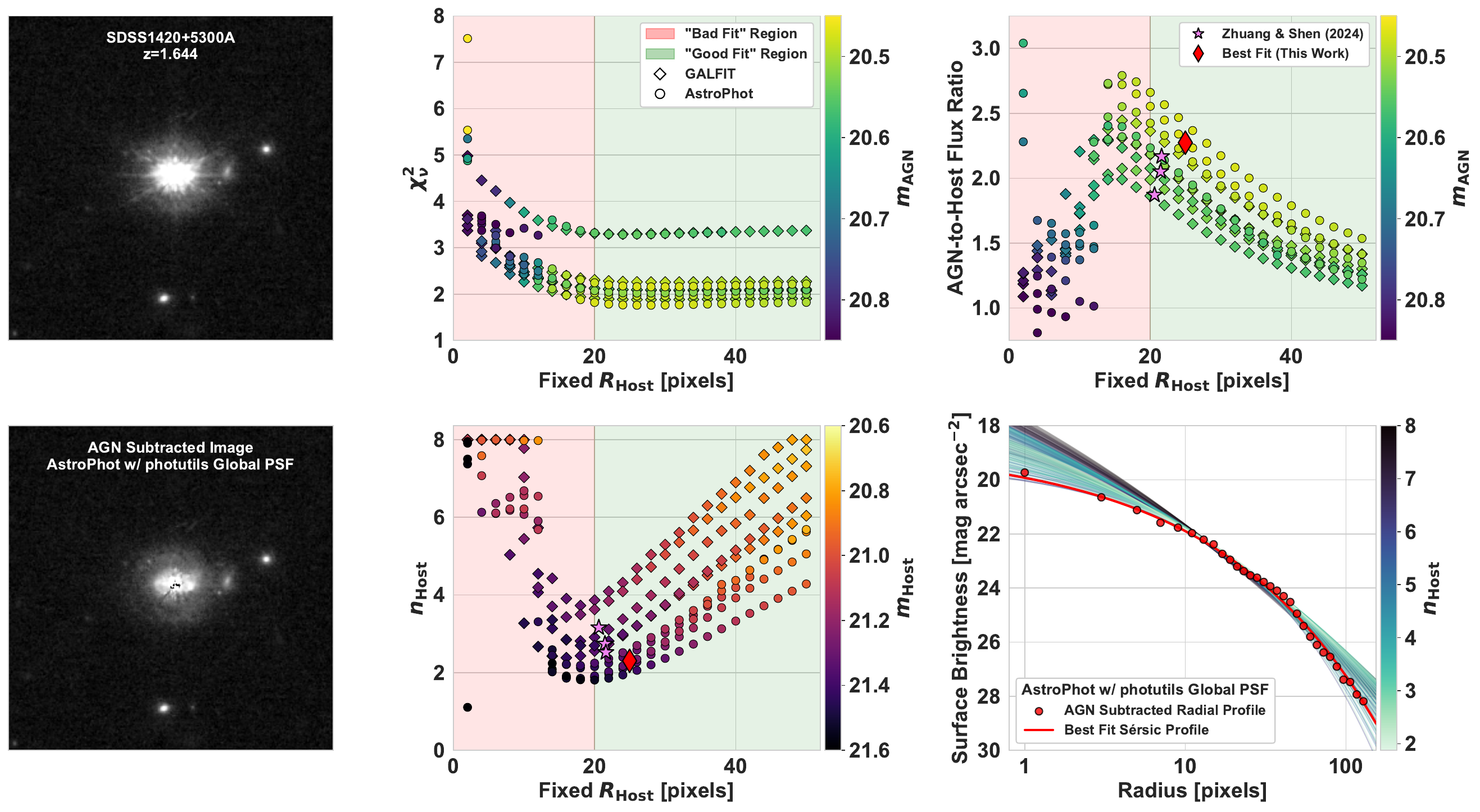}
    \caption{Comparison of the fits which correctly disentangle the AGN from the host galaxy for SDSS1420+5300A. The left column shows the F115W image (top) alongside the AGN subtracted image (bottom) obtained from our \textsc{AstroPhot} best fit using the \textsc{photutils} Global PSF. Each cutout is 255$\times$255 pixels (\SI{7.65}{\arcsec}$\times$\SI{7.65}{\arcsec}). The middle column and upper right panel highlight the parameters obtained for each fit. Each fit has the effective radius held fixed at the given radius, while each other parameter freely varies. The green shaded regions highlight the range within which the $\chi^2_\nu$ indicates a high quality fit, whereas the red region highlights the region where the fits are consistently poor. The pink stars show the best-fit results of \citet{Zhuang2024}. The red diamond shows the best fit results from our \textsc{AstroPhot} \textsc{photutils} Global PSF fit. The lower right panel shows the S\'ersic profiles for the derived host parameters for every fit in the ``Good Fits'' region. The red circles are the median radial profile of the AGN subtracted image shown in the lower left, and the red line shows the best fit of the \textsc{AstroPhot} \textsc{photutils} Global PSF model.}
    \label{fig:Degen_Pretty}
\end{figure*}

Figure \ref{fig:Degen_Pretty} again shows the fits parameters, but removes the groups for which our sources consistently failed, namely the \textsc{PSFEx} Global, Global B, and Epoch 1B PSFs described above. We also remove the $r_e \geq 40$ pixels fits from the \textsc{AstroPhot} \textsc{photutils} Dither 1B fits where the S\'ersic index reaches our constraint. The fits tend to converge to a constant minimum $\chi^2_\nu$ for all fits above $r_e\sim20$ pixels. Thus, we can split our parameter space into a ``Good Fit'' region ($r_e \geq 20$ pixels) and a ``Bad Fit'' region ($r_e < 20$ pixels). Within the ``Good Fit'' region, the AGN magnitude remains relatively constant suggesting that, as long as the minimization algorithm correctly identifies the presence of the host galaxy, the parameters of the host do not significantly affect the AGN magnitude measurement---the choice of PSF plays a more significant role than the fitting software or host parameters.

Looking at the host galaxy parameters in the ``Good Fit'' region, we see a strong, positive correlation between the effective radius, S\'ersic index, and host magnitude, spanning the entire region. This relation covers a significant portion of the available parameter space, over a wide range of S\'ersic indices ($1.8<n<8$) and host magnitude (${\sim}1$ mag). While the host parameters vary significantly, the $\chi^2_\nu$ and residual image do not noticeably change, indicating that the $\chi^2_\nu$ is not a good indicator of quality of fit for our model. Inspecting the residual images for the fits also shows that there is no significant visual indicator suggesting that one fit is ``better'' than another---each fit does an equally sufficient job of modeling the host galaxy outside of the clear non-S\'ersic elements (e.g., spiral arms or bar). Because the AGN magnitude also remains relatively constant, the AGN-to-host flux ratio varies significantly across the parameter space, from ${\sim}1.2$ to ${\sim}2.8$.

For the host galaxy, the \textsc{AstroPhot} and \textsc{Galfit} fits take up completely different regions of the parameter space. As we found generally from examining all of our best fits in Section \ref{sec:results}, \textsc{Galfit} finds consistently higher S\'ersic indices than \textsc{AstroPhot}. Since the S\'ersic index, $n$, is positively correlated with the host magnitude, the higher $n$ value corresponds to a brighter host magnitude and lower AGN-to-host ratio at each radius. In order to find the consistent host magnitude measurement found in Section \ref{sec:results}, the effective radius must be lower for higher values of the S\'ersic index. Figure \ref{fig:Degen_Pretty} also includes the best fit for our \textsc{AstroPhot} \textsc{photutils} Global PSF, alongside the best fit S\'ersic+PSF models from \citet{Zhuang2024}. The \citet{Zhuang2024} points align closely with our \textsc{Galfit} fits. The displayed best fits obtain a relatively consistent host magnitude by balancing the S\'ersic index and effective radius.

By taking each fit in the ``Good Fits'' region and calculating the resulting S\'ersic profile, we can compare how significantly the brightness profile varies over the acceptable parameter space. The lower right panel in Figure \ref{fig:Degen_Pretty} demonstrates this, alongside the AGN-subtracted median profile and associated best-fit S\'ersic profile for our \textsc{AstroPhot} \textsc{photutils} Global PSF fit. We find that the S\'ersic profiles do not vary strongly throughout our parameter space, particularly outside of the central few pixels. Within the central region, the S\'ersic profiles do show fairly significant variation on the order of $2 \ \mathrm{mag \cdot \si{\per\arcsecSpelled\squared}}$. However, this region is significantly contaminated by residuals from the AGN subtraction. In addition, the AGN magnitude remains constant despite the variation in central surface brightness, and so this central region factors very little into the AGN-host galaxy decomposition. In general, the non-S\'ersic elements of the host galaxy cause variations in the radial profile equal to or greater than the distribution of S\'ersic profiles. This implies that the S\'ersic profile, while able to obtain a consistent brightness measurement, is not able to adequately model the radial profile of the galaxy, and the S\'ersic index may not be indicative of the host galaxy's morphology.

\begin{figure*}
    \centering
    \includegraphics[width=\linewidth]{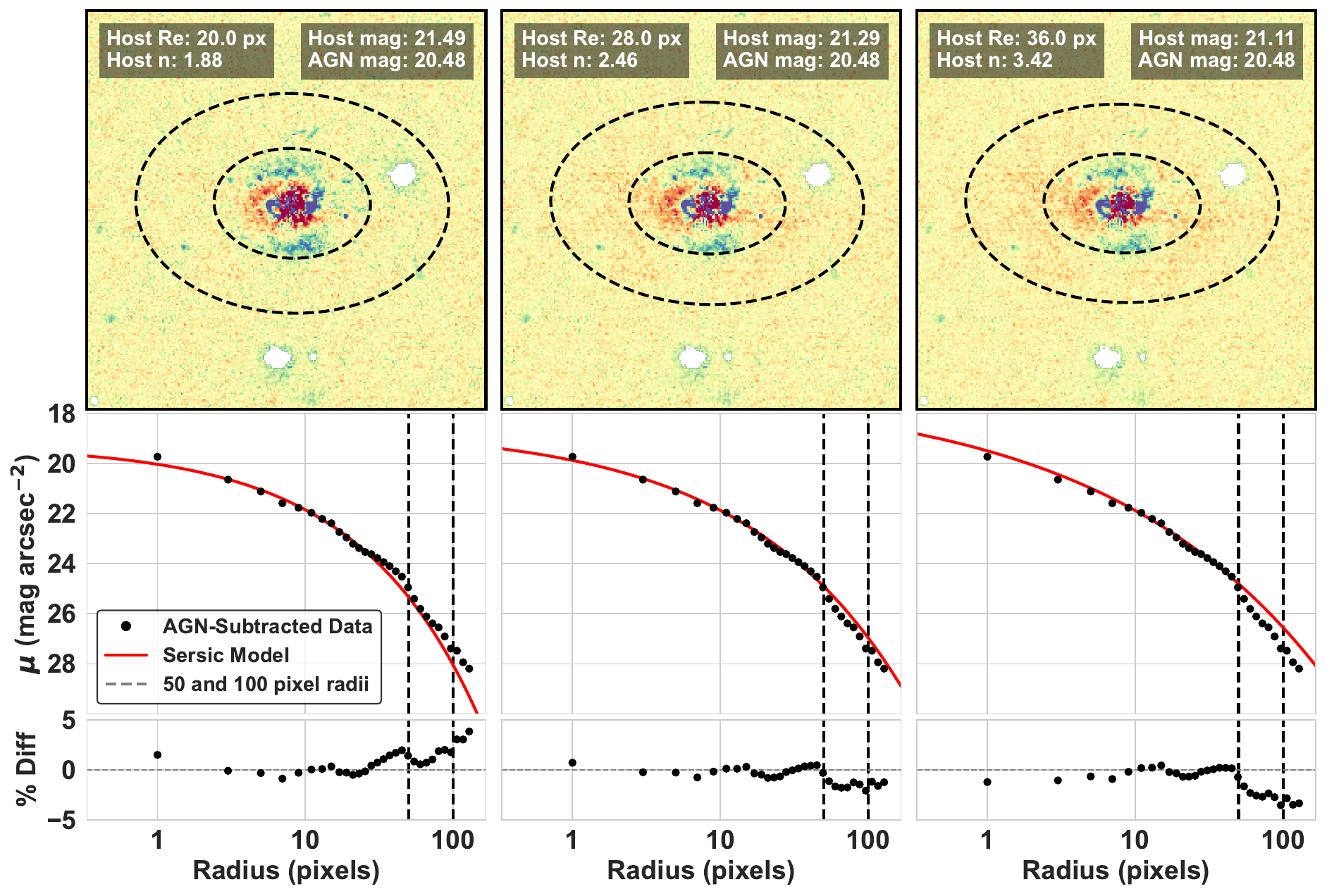}
    \caption{Comparison of three models within the collection of ``Good Fits'' for SDSS1420+5300A shown in Figure \ref{fig:Degen_Pretty}. We include the \textsc{AstroPhot} models with the \textsc{photutils} Global PSF for the effective radii fixed at $R_\mathrm{Host}=20, 28, 36$ pixels (left, middle, and right columns, respectively). The top row of panels shows the residual of the model, including all of the host galaxy, AGN, and a single neighbor component. The blue pixels represent regions where the model underestimated the surface brightness of the source. The red pixels represent regions where the model overestimated the data. The bottom panel shows the 1D profile of the host galaxy (after subtracting the AGN and neighbor) plotted with the S\'ersic profile for the given parameters. The dashed black lines represent the $r=50,100$ pixel radii.}
    \label{fig:Residuals}
\end{figure*}

Figure \ref{fig:Residuals} highlights three ``Good Fits'' of SDSS1420+5300A. All three fits do a reasonably good job of modeling the radial profile of the host galaxy; however, the models span a wide range of effective radii, S\'ersic indices, and AGN-to-host flux ratios. For this example, we find that the $R_\mathrm{Host}=20$ model tends to underestimate the surface brightness of the central region, spiral arms, and extended wings of the source, while the larger radii models tend to model the bar and spiral arms more accurately, but overestimate the central region and wings. While the specifics vary from source to source, it is consistently true that the most important regions for constraining the S\'ersic profile are the central pixels and the extended wings. As previously discussed, the central region is significantly influenced by residual effects from the AGN subtraction, making constraining the profile difficult. The extended wings of the source contribute very little to the overall $\chi^2_\nu$ of the fit---in Figure \ref{fig:Residuals}, which shows minor but clear under- and over-subtraction of the extended wings between models, the $\chi^2_\nu$ of the three models differs by $<1\%$. Thus, the shape of the profile is dominated by the region outside of the FWHM of the PSF, nearer to the effective radius. The larger distribution of profiles shown in Figure \ref{fig:Degen_Pretty} demonstrate significant overlap in this region, leading to a wide range of ``valid'' fit parameters.

Our results highlight a significant problem---if the model parameters are dependent on the software package used, then how can we compare results across different studies? The precise measurement of the S\'ersic index is of little value if a small shift in the effective radius balances the profile, especially for a source with clear non-S\'ersic features, and if different combinations of software probe unique regions of the parameter space. However, the general regime in which the S\'ersic index falls is still absolutely of value; for example, the best fit S\'ersic indices for SDSS1420+5300A typically fall between $1.8<n<3.5$, well within the expected range for a massive, barred spiral galaxy. It is possible that the most effective way to characterize such systems is to combine the S\'ersic+PSF models with other empirical measurements that provide more readily comparable representations of the system.

By starting with S\'ersic+PSF fits, we are able to consistently separate the light of the AGN from its host galaxy. We can then follow up with a number of alternative methods to probe for more consistent information about the host galaxy. \citet{Powell2017} recursively fit inactive galaxies with additional S\'ersic profiles using \textsc{Galfit}---beginning with a single S\'ersic profile, advancing to a two component bulge+disk fit, then applying additional S\'ersic profiles while balancing between quality of fit and minimizing overfitting. Because we can accurately subtract the AGN light, applying this technique to our AGN-subtracted image could more accurately model a complicated galaxy radial profile. Additionally, we could still extract morphological information from the brightest S\'ersic profiles in the model. Similarly, we can utilize the AGN-subtracted images to perform aperture photometry measurements, alongside empirical morphological measurements such as the concentration and asymmetry of the host galaxy.  This methodology will result in far more consistent and non model-dependent measures of the host galaxy morphology.

\section{Conclusion} \label{sec:conclusion}

In this paper, we investigated how applying various PSF-building and surface-brightness fitting software influence the results of the decomposition of an AGN and its host to measure photometry and characterize morphology. We performed fits of 87 sources in the CEERS survey, with redshifts between $0.1<z<4$ and apparent sizes typically within \SIrange{0.1}{3}{\arcsec}. We created 31 groupings of JWST point sources, with which we created PSF models using both \textsc{PSFEx} and \textsc{photutils}. Similarly, we created nine groupings of HST point sources in order to compare the dependence on the choice of point source sample.

The radial profile of the \textsc{PSFEx} PSFs consistently contained an anomalous region near the radius corresponding to the FWHM where the intensity was much lower than any point sources within our sample. The HST \textsc{PSFEx} PSFs displayed no such feature, suggesting that perhaps \textsc{PSFEx} struggled to model the complex core of the JWST PSF. Comparatively, the \textsc{photutils} PSFs better matched the radial profile of the point source population; however, \textsc{photutils} struggled more significantly for groups with fewer point sources, finding consistently broader FWHM and less detailed extended wings.

The \textsc{photutils} PSFs performed more consistently than \textsc{PSFEx}. This is not to say that \textsc{photutils} performs better than \textsc{PSFEx} in every scenario---\citet{Zhuang2024} and \citet{Berman2024} found success in building \textsc{PSFEx} PSFs from significantly larger samples of input point sources. The ability to account for spatial variation across an image is among the major benefits of using \textsc{PSFEx}.  However, extragalactic survey fields typically have too few reference PSFs to take advantage of this feature, and so it is not explored in this work. For situations in which modeling PSF spatial variation is viable, the level of user control and customization available in \textsc{AstroPhot} provide a powerful combination for AGN decomposition.

In exploring the results of surface brightness fits with both \textsc{Galfit} and \textsc{AstroPhot}, the choice of PSF has little influence on the structural parameters of the host galaxy. However, the AGN magnitude does depend on the choice of PSF. The least-populated point source groupings, which have broader FWHMs, result in dimmer AGN magnitudes than the groups with more reference point sources. Between the more highly-populated point source groupings, the AGN measurements are relatively robust. PSFs generated by \textsc{PSFEx} tend to provide fainter AGN magnitudes than the \textsc{photutils} PSFs.

The host parameters are significantly affected by the choice of fitting software, with \textsc{Galfit} and \textsc{AstroPhot} probing two distinct regions of the parameter space---\textsc{Galfit} tends to converge to higher $n$ and lower $r_e$ than \textsc{AstroPhot}. Despite this, the host magnitude measurement is consistent between the two within the region of lowest $\chi^2$, highlighting the issue of degenerate solutions within the S\'ersic profile. The non-S\'ersic components of a host galaxy (e.g., spiral arms or a bar) can make significant contributions to the radial profile of the source; such features limit our ability to model the source with a S\'ersic profile.

We find that the $\chi^2_\nu$ does not adequately indicate the quality of our fits. Significant variation in the host parameters of the galaxy---including values clearly beyond physical feasibility---did not correspond to significant change in the $\chi^2_\nu$. While clear outliers can indicate that a fit is not working as intended, Figures \ref{fig:Degen_RedBlue} and \ref{fig:Degen_Pretty} show that the minimum $\chi^2_\nu$ can correspond to a number of fits, rather than a single ``true'' best fit. Additionally, the most physically reasonable fits may sometimes correspond to a higher $\chi^2_\nu$ value, such as for the \textsc{AstroPhot} \textsc{photutils} Dither 1B fits seen in Figure \ref{fig:Degen_RedBlue}.

We have shown that we can reliably disentangle an AGN from its host galaxy, enabling morphological studies for larger and more diverse samples of AGN. However, between the relatively poor agreement between the S\'ersic profile and the radial profile of host galaxies in the CEERS survey, the consistent disagreement between the host parameters reported by \textsc{Galfit} and \textsc{AstroPhot}, and the poor performance of quality-of-fit indicators such as $\chi^2_\nu$, we suggest that alternative techniques be used in tandem with the traditional S\'ersic+PSF decomposition in order to get consistent, easily comparable measures of the host galaxy morphology.

When performing studies comparable to ours, with a limited number of point sources (${<}100$) to build PSFs and reasonably high redshift AGNs ($0.1<z<4$), we recommend:
\begin{itemize}
\item Using \textsc{photutils} to create  PSFs. With a sufficient sample size, we found our results were much more consistent with \textsc{photutils} than \textsc{PSFEx}. Additionally, the \textsc{photutils} PSFs matched the radial profile of our point sources much more closely than those from \textsc{PSFEx}, and did not feature the consistent anomalous pixels near the FWHM.
\item Prioritizing building PSF models with a sufficient number of point sources (in our case, a minimum of 20). The stochastic variation of a poor PSF model is more significant than the variations due to position or dither pattern.
\item Using \textsc{AstroPhot} to perform  fits. \textsc{AstroPhot} is less sensitive to the choice of PSF and is less likely to have the S\'ersic index fail to converge within typical constraints. Additionally, \textsc{AstroPhot} is open-source and provides a level of user control and customization far beyond what is available through \textsc{Galfit}.
\item Performing model-independent measurements of galaxy morphology, such as calculating the concentration or asymmetry of the host galaxy light after subtracting the AGN.
\end{itemize}

\begin{acknowledgments}

C.D., P.B., and S.C.G. acknowledge the support of the Natural Sciences and Engineering Research Council of Canada (NSERC), [funding reference numbers RGPIN-2024-04039 (P.B.) and RGPIN-2021-04157 (S.C.G)]. Cette recherche a été financée par le Conseil de recherches en sciences naturelles et en génie du Canada (CRSNG), [numéros de référence RGPIN-2024-04039 (P.B.) et RGPIN-2021-04157 (S.C.G)]. C.D. thanks the Ontario Graduate Scholarship (OGS) for their support throughout this project. S.C.G. acknowledges the support of the Canadian Space Agency and a Western Research Leadership Chair Award.

We acknowledge the Anishinaabek, Haudenosaunee, L\=unaap\'eewak and Chonnonton Nations, whose traditional territories are where our research was produced.

\end{acknowledgments}

\facilities{HST (ACS, WFC3), JWST (NIRCam)}
\software{\textsc{AstroPhot} \citep{Stone2023}, \textsc{astropy} \citep{AstropyCollaboration2013, AstropyCollaboration2018, AstropyCollaboration2022}, \textsc{Galfit} \citep{Peng2002, Peng2010}, \textsc{photutils} \citep{Bradley2022}, \textsc{PSFEx} \citep{Bertin2011}, \textsc{SExtractor} \citep{Bertin2011}}

\appendix

\section{Description of Point Source Groupings} \label{sec:app_groups}

Here, we outline the selection criteria for each of the point source groups. The PSFs of both JWST and HST vary with both position on the detector and time \citep{Nardiello2022, Zhuang2024}. We investigate the significance of the effect of these variations on both the PSF model generated using \textsc{photutils} and \textsc{PSFEx}, as well as the influence of the choice of PSF for S\'ersic+PSF fits. Compared to studies such as \citet{Nardiello2022} or \citet{Zhuang2024}, we have a far smaller sample of point sources. Thus, we additionally must balance having spatially and temporally local PSFs with having a significant enough sample size to minimize stochastic variation.

The first grouping is a completely ``Global'' PSF that includes all point sources in our library, regardless of pointing or module. Every group acts as a subset of this Global group. The first subgroups that we use are defined by the signal-to-noise of the point sources. Specifically, we create a ``Global High S/N'' for sources with $\texttt{SNR\_WIN} \geq 50$ and ``Global Low S/N'' for sources with $20 \leq \texttt{SNR\_WIN} < 50$. Next, we assign the ``Global A'' and ``Global B'' groups---these contain all point sources across all 10 pointings, but are split depending on which module imaged the point source. NIRCam is split into two nearly identical modules, named A and B. Each module contains four short wavelength detectors (alongside a single long wavelength detector), labeled 1 through 4. We then further split our Global A/B groups into eight subgroups named ``Global A1--4'' and ``Global B1--4'' based on the specific detector each point source was detected on.

From here, we again return to the parent Global group and split into groups based on the data release. ``Epoch 1'' and ``Epoch 2'' include sources from DR0.5 and DR0.6, respectively. We again break these groups down further based on the module into ``Epoch 1A'', ``Epoch 1B'', ``Epoch 2A'', and ``Epoch 2B''.

We further increase the specificity by splitting the Epoch groups into subgroups based on the dither pattern of the observation. Following \citet{Zhuang2024}, we place CEERS1+2 into a single group called ``Dither 1''. Both of these pointings use a dither pattern tied to the simultaneous MIRI observations, whereas CEERS3+6 are grouped into ``Dither 2'' by virtue of sharing the \texttt{3-POINT-MIRIF770W-WITHNIRCam} dither pattern, causing different step sizes within images of the same NIRCam filter \citep{Bagley2023}.

We analogously split the Epoch 2 groups into two subgroups, ``Dither 3'' and ``Dither 4''. Dither 3 contains the point sources from CEERS4+10, whereas Dither 4 contains CEERS5+CEERS7--9. The primary distinction between these two groups is that Dither 4 has two sets of NIRCam F115W imaging overlaid: one taken as NIRSpec parallels, and one with WFSS imaging. These two sets of data have different readout patterns (\texttt{MEDIUM8} for the NIRSpec parallels, and \texttt{SHALLOW4} for the WFSS imaging). In addition, the NIRSpec-defined imaging used a three point dither, whereas the WFSS pointings used a four point dither. Thus, we have separated these pointings from the rest of Epoch 2.

The last split is to subgroups based on both the dither pattern and the module, creating the final eight groups: ``Dither 1A/B'', ``Dither 2A/B'', ``Dither 3A/B'', and ``Dither 4A/B''. If one is to compare the PSFs in this work to those available from \citet{Zhuang2024}, these final groupings would be the direct comparable. Overall, we have 31 groupings of JWST point sources for which we can create PSFs.

For our HST point sources, we have a smaller set of nine groups: ``Global'', ``Global High S/N'' ($\texttt{SNR\_WIN} \geq 30$), ``Global Low S/N'' ($20 \leq \texttt{SNR\_WIN} < 30$), ``Epoch 1 or 2'', and ``Dither 1--4''. Adopting these groupings, particularly the Epoch and Dither groups, has no inherent physical significance for HST; the Epoch and Dither group are defined based on the JWST imaging independent of the HST imaging and simply exist as a way of providing different groupings of point sources to investigate. We expect that these groupings should provide PSF models more consistent than the JWST groups.

\section{Effect of the PSF on the Fits} \label{sec:app_fits}

\begin{figure*}
    \centering
    \includegraphics[width=\linewidth]{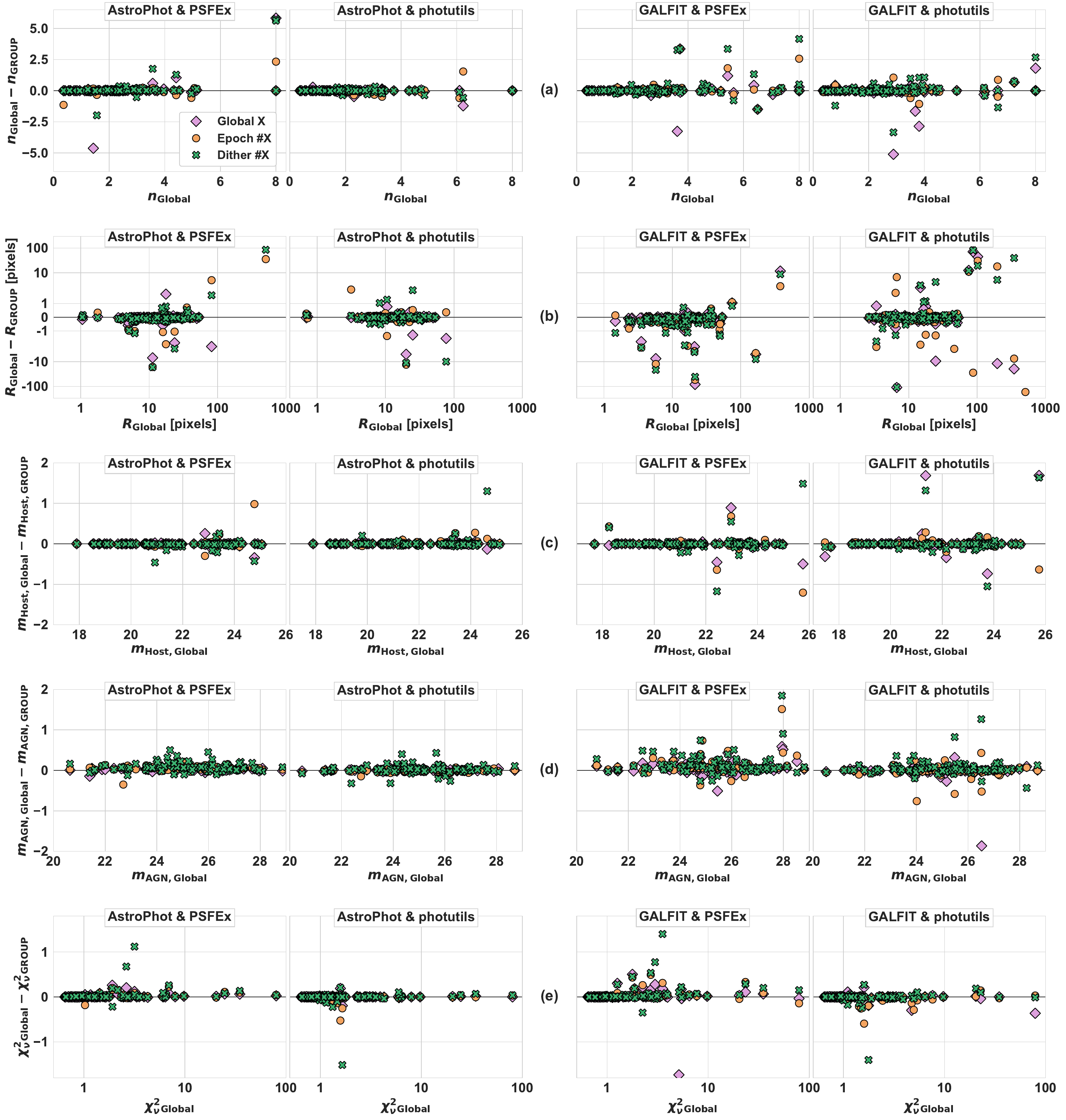}
    \caption{Comparison of the \textsc{AstroPhot} (left) and \textsc{Galfit} (right) fit parameters for the various PSF groupings for both \textsc{PSFEx} and \textsc{photutils}. The panels from top down show: (a) the S\'ersic index, (b) effective radius, (c) host magnitude, (d) AGN magnitude, and (e) $\chi^2_\nu$. Each plot is shown as the difference between the Global PSF group and the Global A/B (pink diamonds), Epoch 1A--2B (orange circles), and Dither 1A--4B (green crosses) groups as a function of the Global group's value. In general, the PSF grouping does not have a strong influence on the results of the fit.}
    \label{fig:Group_vs_Group_All}
\end{figure*}

Here, we investigate how the choice of PSF influences the results of our fits. First, we compare how the fit parameters vary with the choice of point source groupings with \textsc{PSFEx} and \textsc{photutils}. Figure \ref{fig:Group_vs_Group_All} shows the relative difference between the Global PSF grouping and the Global A/B, Epoch 1A-2B, and Dither 1A-4B groupings for each of the main fit parameters from \textsc{AstroPhot} and \textsc{Galfit}; namely, the S\'ersic index, effective radius, host magnitude, AGN magnitude, and the reduced chi-squared value. In general, the choice of PSF is internally consistent within \textsc{PSFEx} or \textsc{photutils} and does not have a significant, consistent influence on the results of the fit for any combination of \textsc{photutils}, \textsc{PSFEx}, \textsc{Galfit}, or \textsc{AstroPhot}.

The \textsc{Galfit} fit parameters do show a marginally higher scatter than \textsc{AstroPhot}, particularly for the effective radius (Figure \ref{fig:Group_vs_Group_All}b) and AGN magnitude (Figure \ref{fig:Group_vs_Group_All}d). There are several outliers in both \textsc{AstroPhot} and \textsc{Galfit}, however these are typically sources that are not particularly well-fit with any grouping. There is a small subset of such sources that are well-modeled by some PSFs but not others---these are generally AGN-dominated sources fit with the poorer (e.g., with lower number of contributing sources) PSF models (e.g., Dither 1A). These poor PSF models are the most significant cause of the scatter in the AGN magnitudes.

\begin{figure*}
    \centering
    \includegraphics[width=0.48\linewidth]{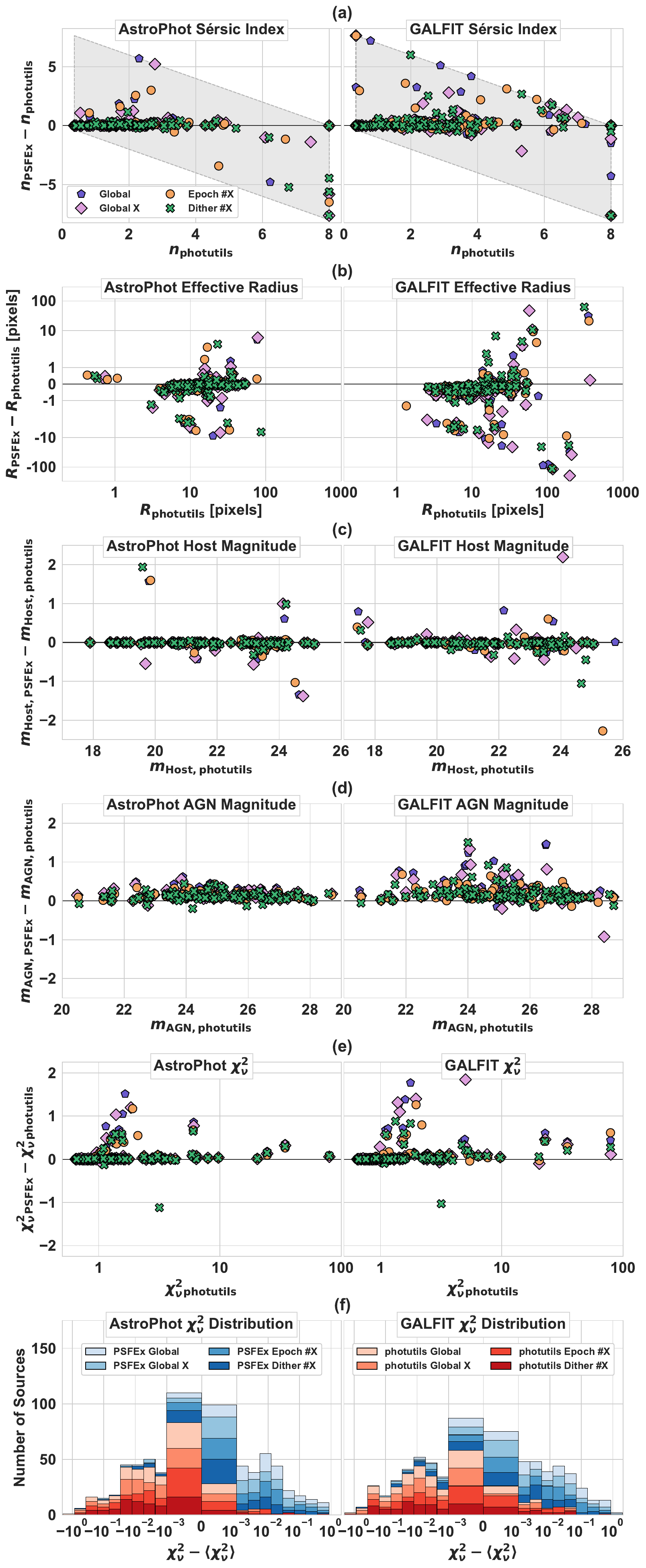}
    \caption{Direct comparison of the fit results for \textsc{photutils} and \textsc{PSFEx} using \textsc{AstroPhot} (left) and \textsc{Galfit} (right). Each row corresponds to a fit parameter, concluding with the distribution of $\chi^2_\nu$ relative to the median of each source. The gray region indicates the region permitted by our constraints. \textsc{Galfit} tends to see more scatter than \textsc{AstroPhot}. The \textsc{PSFEx} PSFs generally have a larger $\chi^2_\nu$ than the \textsc{photutils} PSFs.}
    \label{fig:PSF_vs_PSF}
\end{figure*}

Next, we investigate how the results vary between fits utilizing \textsc{photutils} and \textsc{PSFEx} PSFs. Figure \ref{fig:PSF_vs_PSF} shows the relative difference in the main fit parameters between \textsc{PSFEx} and \textsc{photutils} for both \textsc{AstroPhot} and \textsc{Galfit}. Generally, the agreement is again consistent---however, there is larger scatter than within \textsc{PSFEx} or \textsc{photutils}.

Between PSFs, the host magnitude (Figure \ref{fig:PSF_vs_PSF}c) is by far the most consistent, whereas the S\'ersic index (Figure \ref{fig:PSF_vs_PSF}a), effective radius (Figure \ref{fig:PSF_vs_PSF}b), and AGN magnitude (Figure \ref{fig:PSF_vs_PSF}d) see more significant variance. \textsc{Galfit} is more significantly influenced by the choice of PSF than \textsc{AstroPhot} across all parameters. The scatter is not symmetric---the effective radius is marginally but consistently lower, the S\'ersic index skews higher (particularly for \textsc{Galfit}), and the AGN magnitude tends to be fainter for the \textsc{PSFEx} PSFs.

The $\chi^2_\nu$ of the \textsc{photutils} fits are consistently lower than the \textsc{PSFEx} fits for both \textsc{AstroPhot} and \textsc{Galfit}. This is likely due to the deviation seen in the \textsc{PSFEx} PSFs’ radial profile---if the constructed PSF is not representative of a point source in the image, any fit using such a PSF will result in a higher $\chi^2_\nu$.

\bibliography{references}{}
\bibliographystyle{aasjournalv7}

\end{document}